%% file: o-instantons.tex
\documentclass[12pt,a4paper,leqno]{article}
\input{./packages.sty}
\include{macros}

\title{ADHM in  $8d$, coloured solid partitions
and \\ Donaldson-Thomas invariants on orbifolds}

\date{}
\author[a,b,c]{G.~Bonelli,}
\author[a,b,c]{N.~Fasola,}
\author[a,b,c]{A.~Tanzini,}
\author[a,b,c,d,e,f]{and Y.~Zenkevich}
\affiliation[a]{S.I.S.S.A. -- Scuola Internazionale di Studi Scientifici Avanzati\\ via Bonomea, 265 - 34136 Trieste, Italy}
\affiliation[b]{I.N.F.N. -- Sezione di Trieste}
\affiliation[c]{I.G.A.P. -- Institute for Geometry and Physics\\ via Beirut, 4 -34100 Trieste, Italy}
\affiliation[d]{ITEP -- Institute for Theoretical and Experimental Physics\\Bolshaya Cheremushkinskaya street 25, 117218 Moscow, Russia}
\affiliation[e]{ITMP MSU -- Institute for Theoretical and Mathematical Physics\\Leninskie gory 1, 119991 Moscow, Russia}
\affiliation[f]{MIPT -- Moscow Institute of Physics and Technology\\Institutski per. 9, 141701 Dolgoprudny, Russia}
\emailAdd{bonelli@sissa.it}
\emailAdd{nfasola@sissa.it}
\emailAdd{tanzini@sissa.it}
\emailAdd{yegor.zenkevich@gmail.com}
\abstract{
We study
the moduli space of $SU(4)$ invariant BPS conditions in supersymmetric gauge theory on non-commutative ${\mathbb C}^4$
by means of an ADHM-like quiver construction and we classify the invariant solutions under the natural toric action in terms of solid partitions.
In the orbifold case ${\mathbb C}^4/G$, $G$ being a finite subgroup of $SU(4)$, the classification is given in terms of coloured solid partitions. 
The statistical weight for their counting is defined through the associated 
equivariant cohomological gauge theory. We explicitly compute its partition function on ${\mathbb C}^4$ and 
${\mathbb C}^2\times\left({\mathbb C}^2/{\mathbb Z}_2\right)$
which conjecturally provides the corresponding orbifold Donaldson-Thomas invariants.
}
\begin{document}
\begin{tikzpicture}[remember picture,overlay] 
\draw[draw=none] (13.2,.15) node{ITEP-TH-23/20};
\end{tikzpicture}
\maketitle
\section{Introduction}

The aim of this paper is the study of the moduli space of solutions of an 
eight dimensional analog of the celebrated self-duality equation
$F=\star F$
for the gauge theory curvature in four dimensions \cite{Belavin:1975fg}.
The equation in eight dimensions reads 
\be
F\wedge T= \star F
\label{sd8}
\ee
where $F=dA+A\wedge A$ is the curvature of the gauge bundle,
$T$ is an invariant closed four-form and $\star$ is the Hodge star operator with respect to a given Riemannian structure on the eight dimensional manifold.
Equation \eqref{sd8} was introduced in \cite{Corrigan1983} in 1982.
The very existence of the invariant four-form $T$ restricts the holonomy group of the eight dimensional manifold $X$ to be contained in $\Spin(7)$ \cite{joyce96}. 

We provide the eight dimensional analog of the ADHM  \cite{Atiyah:1978ri} construction for \eqref{sd8} with $U(N)$ gauge group for $X={\mathbb C}^4$
and its discrete Calabi-Yau quotients. As we will discuss, differently from the real four dimensional case,  genuine solutions to the above equation exist only on eight dimensional spaces whose local model is a non-commutative deformation of ${\mathbb C}^4$. The latter is obtained by deforming one of the moment maps
\`a la Nekrasov-Schwarz \cite{Nekrasov:1998ss}.
This further breaks the holonomy group $\Spin(7)\to \Spin(6)\simeq \SU(4)$,
implying that $X$ is a Calabi-Yau fourfold.
The corresponding gauge theory  can be engineered as the low energy limit of a $D(-1)/D7$ system in a stabilising non trivial $B$-field background \cite{witten:2000mf}, aligned along the invariant 2-form associated to the 
deformed moment map.
The more general configuration that we will study includes also a set of 
$\overline{D7}$s which act as a source of matter field/observables, as in \cite{Nekrasov2017,Nekrasov2019}.
By resorting to our higher dimensional ADHM construction, we provide explicit solutions of equation \eqref{sd8} in the abelian case.
Let us notice that 
${\mathbb C}^4$ admits a 
$({\mathbb C}^\ast)^3$ toric action compatible with its (trivial)
Calabi-Yau structure, which naturally lifts to the moduli space of solutions to \eqref{sd8}. 
We describe the invariant solutions supported at the fixed points of this toric action, by making crucial use of the non-commutative deformation, and find that these are classified by solid partitions. These are a four dimensional analog of Young diagrams built with hypercubes accumulating on the corner of ${\mathbb R}_+^4$.
This provides a lift to four complex dimensions of the statistical crystal melting model based on plane partitions,  see \cite{Iqbal:2003ds} for the $\Unitary(1)$ case and \cite{Cirafici:2010bd,Fasola:2020hqa} for $\Unitary(N)$.
All this construction has a natural extension on discrete quotients of 
${\mathbb C}^4$, the fixed points being described in this case by 
coloured solid partitions.\footnote{Let us remark that on (partial) resolutions of  ${\mathbb C}^4$ orbifolds one can also construct abelian instantons whose gauge flux is along non-trivial two cycles.
}

As it is well known, the ADHM construction for four-dimensional instantons  is at the root of an isomorphism with the moduli space of framed torsion free coherent sheaves on $\mathbb{P}^2$
\cite{Donaldson1984}. We provide here an analog
isomorphism between the moduli space of solution of \eqref{sd8} on non-commutative $\mathbb{C}^4$ and the moduli space of framed torsion free coherent sheaves on $\mathbb{P}^4$, 
extending it also to the orbifold case by adapting 
the Kronheimer-Nakajima construction \cite{Kronheimer1990}. In this context the fixed points are described by ideal sheaves on $\mathbb{C}^4$ and its quotients. 

Building on equation \eqref{sd8}, one can construct
\cite{Baulieu:1997jx}
a (semi)Topological Field Theory\footnote{The "semi" refers to the dependence on the four-form $T$, which calibrates the volume of the four-cycles
in the eight dimensional manifold.}, which is indeed a topological twist of the eight dimensional gauge theory describing the $D(-1)/D7$ system 
at low energy 
\cite{Acharya:1997gp,Blau:1997pp}.
This provides the setting for BPS-bound states counting whose mathematical counterpart is given by Donaldson-Thomas theory on four-folds 
\cite{Donaldson1996a}. Let us remark that major progresses have been recently obtained on the compactification of the moduli space of solutions of \eqref{sd8} and a rigorous definition 
of the associated enumerative invariants \cite{Cao2014,Cao2017,Cao2018a,Cao2019a,Oh:2020rnj}.
A natural extension is to consider the theory on $S^1\times X$
computing the Witten index of $D0/D8/\overline{D8}$ bound states whose mathematical counterpart is the lift to K-theory.
On toric manifolds one can study the equivariant extension of the sTQFT so  providing a geometrically motivated statistical weight for counting solid partitions
which describe the fixed points of the gauge theory moduli space \cite{Nekrasov2019}.
Chiral ring observables can be introduced via descent equations as in the four dimensional case \cite{Baulieu:1997jx}. 
Their explicit evaluation on $\mathbb{C}^4$ via equivariant localisation
recently appeared in \cite{Nekrasov2019,Cao2019a}.

Let us remark that M-theory on local four-folds provides a geometric engineering description of supersymmetric gauge theories in three dimensions \cite{Intriligator:2012ue},  
analogously to the much better known case of local three-folds \cite{Hollowood:2003cv} which instead describes five dimensional 
supersymmetric gauge theories. Moreover, interesting classes of $(0,2)$ supersymmetric models in two-dimensions arise from 
$D1$-branes probing toric Calabi-Yau four-fold singularities \cite{Franco:2015tya}.
It is thus interesting to
study the eight dimensional BPS counting problem on some examples of local CY four-fold geometries. 
To this end, 
in this paper we also provide the generalisation of the eight-dimensional ADHM-like quiver to orbifolds 
${\mathbb C}^4/G$, where $G$ is a discrete subgroup of $SU(4)$. The fixed points in this case are classified by coloured solid partitions whose 
statistical weight depends on the representation of $G$.
This boils down to count $G$-coloured 
hypercubes configurations whose colouring rules are dictated by the specific action of  $G$ on ${\mathbb C}^4$.
This provides an eight dimensional analog of instanton counting on four-dimensional ALE spaces \cite{Fucito:2004ry,Bonelli:2011kv,Bonelli:2011jx,Belavin:2011sw,Bonelli:2012ny,Belavin:2012eg,Bruzzo:2013daa,Bruzzo:2014jza}.
As an example,
we explicitly address the associated K-theoretic counting problem on ${\mathbb C}^2\times {\mathbb C}^2/{\mathbb Z}_2$.

There are a number of interesting problems to be addressed. Supersymmetric gauged linear sigma models in two dimensions modelled on the ADHM-like quivers presented in this paper can be studied via localisation technique.
In the sphere case this could possibly shed light on the associated quantum cohomology and its relation with quantum integrable hydrodynamics analogously to the four dimensional ADHM quiver studied in   
\cite{Bonelli:2013rja,Bonelli:2013mma,Bonelli:2014iza}. The torus case would allow to compute the elliptic genus of the eight dimensional ADHM moduli space and as such to provide an elliptic lift of Donaldson-Thomas invariants
on fourfolds analogous to the one studied in 
\cite{Benini:2018hjy} on three-folds.
Finally defect operators can be investigated by a generalisation of the ADHM-like quivers similar to the one studied in \cite{Kanno2011,Nakajima:2011yq,Jeong:2020uxz} for the four-dimensional case
and also by the eight dimensional generalisation of the 
{\it nested instantons} studied in \cite{Bonelli:2019lal,Bonelli:2019het}.

The structure of the paper is the following.
In section \ref{2} we discuss the equivariant extension of the eight-dimensional sTQFT, set up 
the gauge theory framework to count its BPS states 
and describe the relevant equivariant observables.
In section \ref{sec:adhm-construction-8d} we provide the ADHM-like construction solving equation \eqref{sd8}
on ${\mathbb C}^4$ with $B$-field and describe the explicit solutions at the fixed points of the toric action in terms of solid partitions.
In section \ref{4} we extend the $8d$ ADHM quiver construction to orbifolds $\mathbb{C}^4/G$ 
and provide the computation of the equivariant partition function 
on the explicit example of ${\mathbb C}^2\times {\mathbb C}^2/{\mathbb Z}_2$.
In Appendix \ref{sec:warm-up:-standard} we collect basic facts on the classical ADHM construction focusing on the points relevant for its $8d$ generalisation.
Appendix \ref{b} is devoted to the discussion of the relation between $8d$ instantons and sheaf cohomology: in \ref{appendix:framed sheaves} we explain the relation 
between the two relevant moduli spaces via Beilinson's theorem, while in \ref{appendix:orbifold beilinson} we extend it to the orbifold case.

\section{Topological gauge theory in eight dimensions}\label{2}
 
 Let $X$ be a real Riemannian eight dimensional differentiable manifold with a torsion free $\Spin(7)$ structure 
 \cite{Joyce1996}.
This is determined by a real covariantly constant spinor $\psi_+$     
$$T=\psi_+^T\Gamma^{[\wedge 4]}\psi_+\in\Lambda^4(X)$$
        where $\Gamma\equiv\Gamma_{\mu}dx^{\mu}$
        and $\{\Gamma_{\mu}\}_{\mu=1,\ldots,8}$ the $\SO(8)$ $\Gamma$-matrices. We assume $\psi_+$ to be of positive chirality and normalised as 
 $\psi_+^T\psi_+=1$.

On $X$ the vector spaces of p-forms $\Lambda^p(X)$ split in irreducible representations of the holonomy group. 
In particular one has 
$\Lambda^2(X)=\Lambda^2_{7}(X)\oplus\Lambda^2_{21}(X)$.
This split corresponds to the projections on $\omega\in\Lambda^2(X)$ given by
$T\wedge \omega =-3 \star \omega$ and $T\wedge \omega = \star \omega$
respectively.

The Spin-bundles on $X$ are isomorphic to 
$S^+(X)\sim \Lambda^1(X)$ and $S^-(X)\sim\Lambda^0(X)\oplus\Lambda^2_7(X)$.
A supersymmetric gauge theory can be formulated on a $\Spin(7)$ manifold
via a topological twist which uses these isomorphisms \cite{Baulieu:1997jx,Acharya:1997gp,Blau:1997pp}.
The corresponding twisted supersymmetry transformations
of the gauge theory 
can be made equivariant with respect to an isometry $V$ and
read
\bea
&&QA=\Psi
\,,\quad
Q\Psi=\i_V F -iD\Phi
\,,\quad
Q\Phi=\i_V\Psi
\nonumber
\\
&&Q\chi_7=B_7
\,,\quad
QB_7={\cal L}_V\chi_7+i[\Phi,\chi_7]
\,,\quad
Q\eta=\bar\Phi
\,,\quad
Q\bar\Phi=\i_VD\eta+i[\Phi,\eta]\, ,
\label{susyt}
\eea
 where $V$ is any isometry of the $\Spin(7)$ structure, that is $\star L_V=L_V\star$ and $L_V T=0$, where $L_V\equiv \i_Vd+d\i_V$ is the Lie derivative. In \eqref{susyt}, ${\cal L}_V\equiv \i_VD+D\i_V$ is the covariant Lie derivative. Notice that in \eqref{susyt}, $\Psi\in S^+(X)\sim \Lambda^1(X)$ and in 
 the second line $(\eta,\chi_7)$ and $(\bar\Phi,B_7)\in S^-(X)\sim\Lambda^0(X)\oplus\Lambda^2_7(X)$.
 
 The supersymmetric action after the topological twist can be written as a topological term plus a Q-exact one as
 \be
 S=\int_X T\wedge{\rm Tr}\left(F\wedge F\right) +Q\nu\, ,
 \ee
 where
 \be
 \nu=\int_X{\rm Tr}\left[
 i\star\chi_7\wedge F 
 +
  \Psi  \wedge\star(Q\Psi)^\dagger + g^2 \chi_7\wedge\star(Q\chi_7)^\dagger
  + \eta\wedge\star(Q\eta)^\dagger
 \right]\, ,
 \ee
 where $g$ is the Yang-Mills coupling constant which in the topological theory is a gauge fixing parameter.
 In the path-integral, in the $\delta$-gauge $g=0$, the $B_7$ field appears as the Lagrange multiplier
 for the $\Spin(7)$-instanton equation
 \be
 F_7=0\, ,
 \ee
 which is nothing but a rewriting of eq \eqref{sd8}.
 In the following Section we will provide an ADHM-like description of the solutions to the above equation and their moduli space. This will turn out to have positive (virtual) dimension, inducing a $\Unitary(1)_R$-anomaly
 due to the presence of chiral fermionic zero-modes. In order to have a non-vanishing result one has thus to insert non-trivial observables in analogy with the well known case of Donaldson theory in four real dimensions \cite{Witten1988a}.
 The observables are given by non-trivial cohomology classes of the twisted supersymmetry \eqref{susyt}, 
 and can be obtained from an equivariant version of the usual descent equations, see \cite{Bershtein:2015xfa} for the four-dimensional case.
Indeed, the supersymmetry transformations 
\eqref{susyt} can be  rewritten
as the equivariant Bianchi identity for the 
curvature ${\cal F}=F+\Psi+\Phi$ of the universal bundle as \cite{Baulieu:2005bs}
\be
{\cal D F}\equiv
\left(-Q + D +i\iota_V\right)\left(F+\Psi+\Phi\right)=0\,,
\ee
and expanding in the de Rham form degree. 
Picking an ad-invariant polynomial 
${\cal P}$ on the Lie algebra of the gauge group,
we have
\be
Q{\cal P}({\cal F})=\left(d+i\iota_V\right){\cal P}({\cal F})\, ,
\ee
so that one can build the equivariant observables as
intersection of the above with elements of the equivariant cohomology
of the manifold, ${\bf \Omega}\in H^\bullet_V(X)$ as
\be
{\cal O}\left({\bf \Omega},{\cal P}\right)\equiv
\int_X{\bf \Omega}\wedge {\cal P}({\cal F}).
\ee
In the path integral formulation of the gauge theory,
we will actually consider the generating function of the equivariant observables 
through the determinant bundle
\be
{\cal O}_{\rm det}\left({\bf \Omega} \right)\equiv
\int_X{\bf \Omega}\wedge {\rm det}(m {\bf 1} + {\cal F})\, ,
\ee
where $m$ is a generating parameter of the observables. In the calculations of the following Sections, we will consider the K-theoretic uplift of the above, or in other terms the index of the equivariant theory
on $X\times S^1$.
\section{ADHM construction in eight dimensions}
\label{sec:adhm-construction-8d}
In this section we describe an eight dimensional generalisation of the classical ADHM construction in four dimensions and show that it describes the moduli space of solutions to \eqref{sd8}.
For the sake of completeness we recall in Appendix \ref{sec:warm-up:-standard} the four-dimensional ADHM construction and highlight few aspects of the latter which the reader could find useful to follow the eight dimensional generalisation.

Let us start by fixing the spinorial notation to write 
equation\eqref{sd8}
and the ADHM representation of its solutions. The $\Cliff(8)$
gamma matrices can be chosen as $16 \times 16$ real matrices of the
form
\begin{equation}
  \label{eq:25}
  \Gamma^{\mu} = \left(
    \begin{array}{cc}
      0 & \Sigma^{\mu}\\
      \bar{\Sigma}^{\mu} & 0
    \end{array}
\right),
\end{equation}
where $\Sigma^0 = \bar{\Sigma}^0 = 1_{8 \times 8}$ and $\Sigma^i = -
\bar{\Sigma}^i$ for $i = 1, \ldots, 7$. The latter are real
antisymmetric matrices (they are in fact $\sqrt{-1}$ times purely
imaginary $\Cliff(7)$ gamma-matrices). Let $S_{\pm}$ denote
eight-dimensional real irreducible Majorana-Weyl spinor
representations of $\Spin(8)$ of positive and negative chirality
respectively. Since the representations $S_{\pm}$ are real, the
matrices of $\Spin(8)$ generators
\begin{equation}
  \label{eq:26}
  \Gamma^{\mu \nu} = \frac{1}{2}[\Gamma^{\mu}, \Gamma^{\nu}] = \frac{1}{2}\left(
    \begin{array}{cc}
      \Sigma^{[\mu} \bar{\Sigma}^{\nu]} & 0\\
      0 & \bar{\Sigma}^{[\mu} \Sigma^{\nu]}
    \end{array}
\right)
\end{equation}
are real and antisymmetric and so are the $8\times 8$ blocks
$\Sigma^{\mu \nu} = \frac{1}{2}\Sigma^{[\mu} \bar{\Sigma}^{\nu]}$ and
$\bar{\Sigma}^{\mu \nu} = \frac{1}{2} \bar{\Sigma}^{[\mu}
\Sigma^{\nu]}$. Formulated differently, we have (cf.~\eqref{eq:16}) an
isomorphism of three $\Spin(8)$ representations spaces, each
represented by real and antisymmetric $8\times 8$ matrices: 
\begin{equation}
  \label{eq:27}
  \Lambda^2 S_{+} = \Lambda^2 S_{-} = \Lambda^2 \mathbb{R}^8 = \mathrm{adj}_{SO(8)}.
\end{equation}
The triality of $\Spin(8)$ permutes $S_{+}$, $S_{-}$ and
$\mathbb{R}^8$, which is the defining representation of $SO(8)$. Notice that
each of the two sets of $28$ matrices
$(\Sigma^{\mu\nu})_{\alpha}{^{\beta}}$ -- or
$(\bar{\Sigma}^{\mu\nu})_{\alpha}{^{\beta}}$ -- form a basis in the space
of real antisymmetric matrices\footnote{A third  basis of
  matrices corresponding to the representation $\Lambda^2
  \mathbb{R}^8$ is given by $(\delta^{\mu}_{\alpha}
  \delta^{\nu}_{\beta}- \delta^{\mu}_{\beta}
  \delta^{\nu}_{\alpha})$.}. Due to this fact, the following Fierz
identities hold:
\begin{equation}
  \label{eq:30}
  (\Sigma^{\mu\nu})_{\alpha\beta}
  (\Sigma^{\mu\nu})_{\gamma\delta} = (\bar{\Sigma}^{\mu\nu})_{\alpha\beta}
  (\bar{\Sigma}^{\mu\nu})_{\gamma\delta} = -8(\delta_{\alpha \delta}
  \delta_{\beta \gamma} - \delta_{\beta \delta}
  \delta_{\alpha \gamma }).
\end{equation}
The coefficient in the l.h.s.\ of Eq.~\eqref{eq:30} can be obtained by
contracting $\beta$ and $\gamma$ indices and using the relations of
the Clifford algebra to get $\Gamma^{\mu\nu} \Gamma^{\mu\nu} = - 56
\cdot 1_{16 \times 16}$.

\subsection{Eight-dimensional equations}
\label{sec:self-dual-equat}
There is no way to formulate a first order equation for
$F_{\mu\nu}$ in $8d$ in an $\SO(8)$-invariant manner. Another way to
formulate this is to say that there is no $\SO(8)$-invariant four-index
tensor $T^{\mu\nu\lambda\rho}$, which can be used to write
\begin{equation}
  \label{eq:28}
  \lambda F_{\mu \nu} = \frac{1}{2} T^{\mu \nu \lambda \rho} F_{\lambda \rho}.
\end{equation}
with $\lambda$ being some eigenvalue. However, if we make a choice of
a constant spinor on $\mathbb{R}^8$ we can build from it a tensor
$T^{\mu\nu\lambda\rho}$ invariant under $\Spin(7) \subset
\Spin(8)$. This is the largest possible symmetry subgroup which can be
preserved by equations of the form~\eqref{eq:28} in eight space-time
dimensions. For this construction let us fix $\psi_{+} \in S_{+}$ such
that $\psi_{+}^T \psi_{+} = 1$ and write
\begin{equation}
  \label{eq:29}
  T^{\mu\nu\lambda\rho} = \psi_{+}^T \Gamma^{\mu\nu\lambda\rho} \psi_{+},
\end{equation}
where
\begin{equation}
  \label{eq:31}
  \Gamma^{\mu\nu\lambda\rho} = \frac{1}{4!} \Gamma^{[\mu} \Gamma^{\nu}
  \Gamma^{\lambda} \Gamma^{\rho]}
\end{equation}
The tensor $T^{\mu\nu\lambda\rho}$ then satisfies the $8d$
self-duality equation\footnote{We could have started with a negative
  chirality spinor $\psi_{-}$ corresponding to anti-self-dual
  $T^{\mu\nu\lambda\rho}$. The resulting construction is isomorphic
  due to the triality of $SO(8)$.}
\begin{equation}
  \label{eq:32}
  T^{\mu\nu\lambda\rho} =
   \epsilon^{\mu\nu\lambda\rho\alpha\beta\gamma\delta} T^{\alpha\beta\gamma\delta},
\end{equation}
since
\begin{equation}
  \label{eq:20}
   \epsilon^{\mu\nu\lambda\rho\alpha\beta\gamma\delta}
   \Gamma^{\alpha\beta\gamma\delta} = \Gamma^{\mu\nu\lambda\rho} \Gamma^9.
\end{equation}
For definiteness we take\footnote{In our conventions the spinor
  indices run from $0$ to $7$ similarly to the indices of the
  $\mathbb{R}^8$ vectors.}  $\psi_{+}^{\alpha} =
\delta^{\alpha}_0$. The choice of the spinor $\psi_{+}$ allows us to
split $S_{+}$ into the one-dimensional subspace proportional to
$\psi_{+}$ and the seven-dimensional orthogonal complement which we
call $\tilde{S}_{+}$. In a group-theoretical language this corresponds
to the splitting of the representation $S_{+}$ into $\mathbf{1} \oplus
\mathbf{7}$ under the subgroup $\Spin(7) \subset \Spin(8)$.

The irreducible two-form representation $\Lambda^2 \mathbb{R}^8 =
\mathbf{28}$ of $\Spin(8)$ splits into a sum of two irreps $\mathbf{7}
\oplus \mathbf{21}$ of $\Spin(7)$. These two irreps correspond to two
different eigenvalues $\lambda=1$ and $\lambda=-3$ respectively in the first order field
equations~(\ref{eq:28}). In this way the splitting allows us to write
two different $\Spin(7)$-invariant conditions\footnote{These two
  conditions may be viewed as analogues of the self-duality and
  anti-self-duality conditions in $4d$. However, the latter are more
  similar to choosing the opposite chirality spinor $\psi_{-}$ instead
  of $\psi_{+}$.} on the field strength $F_{\mu\nu}$. The conditions
correspond to the vanishing of the component of $F_{\mu \nu}$ lying in
one of the two irreps of $\Spin(7)$, or equivalently to the
eigenspaces corresponding to two different eigenvalues in
Eq.~(\ref{eq:28}). The choice $\lambda=1$ gives Eq.~\eqref{sd8}.


As discussed in \cite{Corrigan1983} \eqref{sd8} reads then in spinorial form as
\begin{equation}
  \label{eq:39}
   (\Sigma^{\mu \nu})_{\alpha \beta} \psi_{+}^\beta F_{\mu \nu}
  = (\Sigma^{\mu \nu})_{\alpha 0} F_{\mu \nu} = 0,
\end{equation}
which imposes $7 N^2$ equations, so that -- together with $N^2$ gauge
fixing conditions for the gauge group $\Unitary(N)$ -- eliminate all
functional degrees of freedom from $A_{\mu}(x)$ and therefore a finite
dimensional moduli space $\mathcal{M}_{N,k}$ of solutions
remains. Similarly to the $4d$ case, one can view $\mathcal{M}_{N,k}$
as an \emph{octonionic} quotient of the space of connections
$\mathcal{A}$ by the gauge group $\mathcal{G}$. Indeed we can
introduce seven natural symplectic forms $\omega^{(A)}_{\mu \nu} =
(\Sigma^{\mu \nu})_{A 0}$ on $\mathbb{R}^8$ and use them to write seven
symplectic structures on $\mathcal{A}$:
\begin{equation}
  \label{eq:42}
  \Omega_A[\delta_1 A_{\mu}(x), \delta_2 A_{\mu}(y)] =
  \int_{\mathbb{R}^8}  T \wedge \omega^{(A)} \wedge \delta_1 A(x) \wedge \delta_2
  A(x),
\end{equation}
where $T$ denotes the four-form with components $T^{\mu \nu \lambda
  \rho}$. Then the $7N^2$ conditions \eqref{eq:39} correspond to the
vanishing of the seven moment maps
\begin{equation}
  \label{eq:48}
  \mu^A [\phi(x)] = \int_{\mathbb{R}^8} T \wedge \omega^{(A)} \wedge \tr \phi(x) F,
\end{equation}
and we have the "octonionic" quotient 
\begin{equation}
  \label{eq:41}
  \mathcal{M}_{N,k} = \mathcal{A}/\!/\!/\!/\!/\!/\!/\mathcal{G}.
\end{equation}

\subsection{Derrick's theorem and noncommutativity}
\label{sec:dereks-theor-nonc}
Any solution of the first order equations~\eqref{eq:28} is
automatically a solution of the $8d$ Yang-Mills equations. Indeed,
\begin{equation}
  \label{eq:57}
  D_{\mu} F_{\mu \nu} = \frac{1}{2} T_{\mu \nu \lambda \rho}
 D_{\mu} F_{\lambda \rho} = \frac{1}{2} T_{\mu \nu \lambda \rho}
 D_{[\mu} F_{\lambda \rho]}=0
\end{equation}
where we have used the Bianchi identity for $F_{\mu \nu}$ and the fact
that $T_{\mu \nu \lambda \rho}$ is totally antisymmetric. 
We are looking for localized solutions, {\it i.e.} for those sufficiently rapidly
decaying at infinity in order to have finite action. Then, the
solutions of~\eqref{eq:28}, if any exist, should be true
extrema of the Yang-Mills action. The well-known Derrick's theorem
states that in dimensions greater than four no such localized solution
are possible. The idea of the proof is to provide for any non-singular
field configuration a continuous family of configurations with lower
action, so that no true minimum can exist. The family of
configurations is obtained by scaling the initial configuration into
smaller and smaller volume. A simple power counting then shows that
the action on the scaled down configuration is lower. 

Thus, classically, the moduli space $\mathcal{M}_{N,k}$ of solutions
is empty. However, there is a natural way to deform the problem to get
a non-empty space of solutions by introducing noncommutativity. The
commutation relations for the coordinates are similar to those of the
$4d$ case (cf.~(\ref{eq:54})):
\begin{equation}
  \label{eq:58}
  [x^{\mu}, x^{\nu}] = i \zeta (\omega^{-1})^{\mu \nu},
\end{equation}
where $\zeta$ is a real parameter and $\omega_{\mu \nu}$ is a
non-degenerate constant 2-form on $\mathbb{R}^8$. In this case
Derrick's theorem doesn't apply, since the coordinates cannot be
rescaled without affecting their commutation
relations~\eqref{eq:58}. To put it another way, the non-commutativity
introduces an additional fundamental scale $\sqrt{\zeta}$ into the
problem, which puts a limit on how much one can scale down localized
field configurations. So, even when no classical non-singular
solutions to field equations exist, additional solutions to the
non-commutative version of the problem having typical size
$\sqrt{\zeta}$ may appear. This is exactly the situation we have with
Eq.~\eqref{eq:28}, where there is a finite-dimensional moduli
space of solutions after the non-commutative deformation. This will be
our definition of $\mathcal{M}_{N,k}$.
In the next section we will describe this moduli space using an
analogue of the ADHM construction.

\subsection{\texorpdfstring{$8d$ ADHM construction}{8d ADHM construction}}
\label{sec:8d-adhm-construction}
The ADHM equations for the $8d$ instantons were written in 
\cite{Nekrasov2017, Nekrasov2019}. They correspond to bound states of $k$
D0 and $N$ D8 branes in a suitable $B$-field background. As in
sec.~\ref{sec:non-comm-deform}, the $B$-field introduces
non-commutativity into the $8d$ gauge theory. Also, as we have
explained in sec.~\ref{sec:dereks-theor-nonc} it allows for the very 
existence of solutions to the first order equations~\eqref{eq:28},
corresponding to stable low-energy bound states of D-branes.

To introduce the non-commutativity we pick one of the seven complex
structures on $\mathbb{R}^8$, or correspondingly one of the symplectic
forms $\omega^{(A)}_{\mu \nu}$ to represent the K\"ahler form. For
definiteness we choose $\omega^{(1)}_{\mu \nu}$ and denote the
projection of the two-form $B$ on $\omega^{(1)}_{\mu \nu}$ by $\zeta$.

The choice of the complex structure breaks down the $\Spin(7)$
symmetry of the seven first order equations~(\ref{eq:39}) further to
$\Spin(6) = \SU(4)$. Equivalently we can say that by choosing a complex
structure we introduce one more fixed chiral spinor $\chi_{+}^{\alpha}
= \delta^{\alpha}_1$ into our theory (which corresponds to the index
$1$ in $\omega^{(1)}_{\mu \nu}$). The seven equations~(\ref{eq:39}),
transforming as a $\Spin(7)$ spinor, split into an $\SU(4)$ singlet
(corresponding to the component in the direction of $\chi_{+}$) and
further six equations lying in the representation
\begin{equation}
  \label{eq:47}
  \mathbf{6} = \mathbb{R}^6_{SO(6)} =
  (\Omega^{(2,0)}\oplus \Omega^{(0,2)} \mathbb{C}^4)_{+},
\end{equation}
of complex two-forms obeying $\alpha_{z_i z_j} = \epsilon_{z_i z_j z_k
  z_l} \alpha_{\bar{z}_k \bar{z}_l}$.

Unlike in the $4d$ case the $B$-field not only adds a constant to the
value of one of the seven moment maps $\mu^A$, but introduces new
degrees of freedom corresponding to the rectangular matrix $I$, which
only appears in one of the moment map equations.

Somewhat similarly to the $4d$ case we introduce a $(8k + N) \times
8k$ matrix $\Delta(x)$, which can be written as\footnote{Notice that
  the combination $\psi_{+} + i\chi_{+}$ transforms in the complex
  one-dimensional Weyl spinor representation of the $\Spin(2) = U(1)$
  part of $\Spin(6) \times \Spin(2) \subset \Spin(8)$.}
\begin{equation}
  \label{eq:46}
  \Delta(x) = \left(
    \begin{array}{c}
      (B_{\mu}-x_{\mu} 1_{k \times k}) \otimes \bar{\Sigma}^{\mu}\\
            I^{\dag} \otimes (\psi_{+}^{\dag} + i \chi_{+}^{\dag})
    \end{array}
\right),
\end{equation}
where $B_{\mu}$ are eight Hermitian $k \times k$ matrices,
$\Sigma^{\mu}$ are defined in Eq.~(\ref{eq:25}), $I^{\dag}$ is an $N
\times k$ matrix.

We will be looking for solutions
of
\begin{equation}
  \label{eq:50}
  \Delta^{\dag}(x) U(x) = 0,
\end{equation}
with $\Delta$ satisfying certain moment map conditions. However, differently from the $4d$ case,
these conditions do not imply that  $\Delta^{\dag} \Delta =
1_{8 \times 8} \otimes f^{-1}_{k \times k}$
\footnote{Indeed, since $(\Sigma^{\mu \nu})_{\alpha
  \beta}$ is a complete basis of real antisymmetric $8\times 8$
matrices,
\begin{multline}\nonumber
  \Delta^{\dag}(x) \Delta(x) = I I^{\dag} \otimes (\psi_{+}+i
  \chi_{+}) (\psi_{+}^T - i\chi_{+}^T) +    \\
  + \frac{1}{2}
  ([B_{\mu},B_{\nu}] + i \zeta \omega_{\mu \nu}^{(1)} 1_{k\times k})\otimes \Sigma^{\mu
    \nu} + (B_{\mu} - x_{\mu} 1_{k\times k})(B_{\mu} - x_{\mu}
  1_{k\times k}) \otimes 1_{8\times 8} = 1_{8 \times 8} \otimes
  f^{-1}_{k \times k}
\end{multline}
implies (recall that $\omega_{\mu \nu}^{(1)} = (\Sigma^{\mu \nu})_{01}$)
\begin{equation}
  \label{eq:52}
  [B_{\mu}, B_{\nu}] + i \zeta \omega_{\mu \nu}^{(1)} \otimes 1_{k\times
  k} +  i \omega_{\mu \nu}^{(1)} I I^{\dag}  = 0,
\end{equation}
which gives $28$ matrix equations, instead of just seven.}.
Indeed, to solve Eq.\eqref{eq:39}, it is enough to impose 
\begin{equation}
  \label{eq:53}
  \Delta^{\dag} \Delta \psi_{+} =  \psi_{+} \otimes f^{-1}_{k \times k}.
\end{equation}
In our convention for $\psi_{+}$ and $\chi_{+}$ Eq.~\eqref{eq:53} has
explicit components
\begin{equation}
  \label{eq:60}
  (\Delta^{\dag} \Delta)_{A0} = 0, \qquad A=1,\ldots,7,
\end{equation}
or more explicitly
\begin{equation}
  \label{eq:61}
 i (\omega^{(A)\, -1})_{\mu \nu} [B_{\mu}, B_{\nu}] + \delta^{A,1} 
 II^{\dag} = \zeta \delta^{A,1}, \qquad A=1,\ldots,7.
\end{equation}

\subsection{Formulation in complex coordinates}
\label{sec:complex-language}
By explicitely using the complex structure, we can rewrite Eq.\eqref{eq:61}
in the form given in \cite{Nekrasov2017}. The $8d$ ADHM data contains four
complex $k \times k$ matrices $B_a$ and a complex $k \times N$ matrix
$I$ and the equations read
\begin{align}
  \label{eq:49}
  \sum_{a=1}^4 [B_a, B_a^{\dag}] + I I^{\dag}&= \zeta 1_{k \times k}, \\
  [B_a, B_b] - \frac{1}{2} \epsilon_{abcd} [B^{\dag}_c, B^{\dag}_d]&=
  0. \label{eq:65}
\end{align}

The matrix $\Delta^{\dag}$ defined in Eq.~(\ref{eq:46}) can be written
quite explicitly in complex coordinates. Indeed, the main ingredient
of $\Delta^{\dag}$ is the matrix $x_{\mu} \Sigma^{\mu}: S_{-} \to
S_{+}$, which acts from one Majorana-Weyl representation of $\Spin(8)$
to another. Under the $\SU(4)$ subgroup the spinor representations
$S_{\pm}$ split into sums of even and odd parts of the exterior
algebra:
\begin{align}
  \label{eq:101}
  S_{+} = \mathbf{1} \oplus \mathbf{6} \oplus \mathbf{1} &= \Lambda^0
  \mathbb{C}^4 \oplus \Lambda^2 \mathbb{C}^4\oplus \Lambda^4
  \mathbb{C}^4,\\
  S_{-} = \mathbf{4} \oplus \bar{\mathbf{4}} &= \Lambda^1 \mathbb{C}^4
  \oplus \Lambda^3 \mathbb{C}^4.
\end{align}
The matrix $x_{\mu} \Sigma^{\mu}$ then acts on the exterior powers as
an operator
\begin{equation}
  \label{eq:102}
  \imath_{z_a \partial_{z_a}} +
\bar{z}_a dz_a,
\end{equation}
where $\imath_{z_a \partial_{z_a}}$ is the substitution of the vector
field $z_a \partial_{z_a}$. In this
way we get:
\begin{equation}
  \label{eq:103}
  \Delta^{\dag} = \left(
    \begin{array}{cccc|cccc|c}
      b_1 & b_2 & b_3 & b_4 & 0& 0& 0& 0& I\\
      \hline
      b_2^{\dag} & - b_1^{\dag} & 0 & 0 & 0 & 0 & b_4 & - b_3 & 0\\
      b_3^{\dag} & 0 &- b_1^{\dag}  & 0 & 0 & -b_4 &0 &  b_2 & 0\\
      b_4^{\dag} & 0 & 0 & - b_1^{\dag} & 0 & b_3 & -b_2 &0 &   0\\
      0 & b_3^{\dag} &   - b_2^{\dag} & 0 &  b_4 &0 &0 & -b_1 &   0\\
      0 & - b_4^{\dag} & 0 &   - b_2^{\dag}  &  b_3 &0 &-b_1 & 0 &   0\\
      0 & 0 & b_4^{\dag}  &   - b_3^{\dag}  &  b_2 &-b_1&0  & 0 &   0\\
      \hline
       0& 0& 0& 0& b_1^{\dag} & b_2^{\dag} & b_3^{\dag} & b_4^{\dag} & 0
    \end{array}
\right),
\end{equation}
where we have abbreviated $b_a = B_a - z_a$. Notice that unlike in the
four-dimensional case both representations $S_{\pm}$ are real, i.e.\ they admit outer automorphisms which square
to one. Explicitly these automorphisms are given by the matrices
\begin{equation}
  \label{eq:104}
  \tau_{+} = \left(
    \begin{array}{c|cccccc|c}
      0 & 0 & 0 & 0 & 0& 0& 0& 1\\
      \hline
      0 & 0 & 0 & 0 & 0& 0& 1& 0\\
      0 & 0 & 0 & 0 & 0& 1& 0& 0\\
      0 & 0 & 0 & 0 & 1& 0& 0& 0\\
      0 & 0 & 0 & 1 & 0& 0& 0& 0\\
      0 & 0 & 1 & 0 & 0& 0& 0& 0\\
      0 & 1 & 0 & 0 & 0& 0& 0& 0\\
      \hline
      1 & 0 & 0 & 0 & 0& 0& 0& 0
    \end{array}
\right), \qquad \tau_{-} = \left(
  \begin{array}{c|c}
    0 & 1_{4 \times 4}\\
    \hline
    1_{4\times 4} & 0
  \end{array}
\right).
\end{equation}
The square part of the matrix $\Delta$ commutes with the automorphisms in the sense that
\begin{equation}
  \label{eq:105}
  \tau_+ \Delta_{\mathrm{square}}^{\dag} \tau_- = \Delta_{\mathrm{square}}^{*}.
\end{equation}

\subsection{Matrix formulation}
\label{sec:nonc-inst-equat}
After choosing 
the
non-commutative deformation~(\ref{eq:58}), the original instanton
equations~(\ref{eq:39}) become:
\begin{gather}
  F_{ab} = \frac{1}{2} \epsilon_{abcd} F_{\bar{c} \bar{d}},\label{eq:68}\\
  \sum_{a=1}^4 F_{a \bar{a}} = 0.\label{eq:69}
\end{gather}

Let us rewrite these equations in terms of the matrix variables analogously to the noncommutative $4d$ case 
recalled in sec.~\ref{sec:matrix-form-non}. Plugging the $Z_a$ variables in the instanton
equations~(\ref{eq:68}),~(\ref{eq:69}) we get:
\begin{gather}
  [Z_a, Z_b] = \frac{1}{2}\epsilon_{abcd} [Z^{\dag}_{\bar{c}},
  Z^{\dag}_{\bar{d}}],\label{eq:71}\\
  \sum_{a=1}^4 [Z^{\dag}_{\bar{a}}, Z_a ] = 2 \zeta. \label{eq:70}
\end{gather}
Again the nontrivial r.h.s.\ in Eq.~(\ref{eq:70}) arises because of
the noncommutativity of $z_a$ and $z^{\dag}_{\bar{a}}$. The vacuum
solution of Eqs.~(\ref{eq:71}),~(\ref{eq:70}) is given by
\begin{equation}
  \label{eq:72}
  Z^a = z^a,
\end{equation}
and corresponds to vanishing gauge potential $A_a$. Let us now discuss the simplest non trivial solutions.

\subsection{\texorpdfstring{$\Unitary(1)$ one-instanton}{U(1) one-instanton}}
\label{sec:1-inst-solut}
Nontrivial solutions to Eqs.~(\ref{eq:71}),~(\ref{eq:70})
correspond to nontrivial ideals in the ring of polynomials in four
variables. Let us consider the simplest solution corresponding to a
single abelian instanton sitting at the origin of $\mathbb{C}^4$. In
this case (cf. also \cite[Eq.~(3.18)]{Nekrasov:2002kc})
\begin{equation}
  \label{eq:73}
  Z_a = S_{[[[1]]]} z_a f_{[[[1]]]}(N) S_{[[[1]]]}^{\dag},
\end{equation}
where $N = \sum_{a=1}^4 a^{\dag}_{\bar{a}} a_a$ and
\begin{equation}
  \label{eq:75}
  f_{[[[1]]]}(N) = \left( 1 - \frac{24}{N(N+1)(N+2)(N+3)} \right)^{\frac{1}{2}},
\end{equation}
and $S_{[[[1]]]}$ is the partial isometry of the Hilbert space
(which is $ \mathcal{H} = \mathbb{C}[z_1,z_2,z_3,z_4]$), satisfying
\begin{equation}
  \label{eq:74}
  S_{[[[1]]]} S_{[[[1]]]}^{\dag} = 1, \qquad S_{[[[1]]]}^{\dag} S_{[[[1]]]} = 1 - | 0,0,0,0 \rangle
  \langle 0,0,0,0| = P_{\mathcal{H} \backslash \{ |0,0,0,0\rangle \}}.
\end{equation}
Notice that $N$ in the denominator of $f_{[[[1]]]}(N)$ is never zero, because the state $|0,0,0,0\rangle$
is projected out by the partial isometry  .

\subsection{\texorpdfstring{$\Unitary(1)$ multi-instanton}{U(1) multi-instanton}}
\label{sec:multi-inst-solut}
By taking the square of equations~(\ref{eq:71})
we deduce that the operators $Z_a$ commute with each other:
\begin{equation}
  \label{eq:76}
  [Z_a, Z_b] = 0.
\end{equation}
As we have noticed above the multi-instanton solutions correspond to
ideals in the ring $\mathcal{H} = \mathbb{C}[z_1,z_2,z_3,z_4]$. Having
such an ideal $\mathcal{I}$, we define a partial isometry
$S_{\mathcal{I}}$, which satisfies
\begin{equation}
  \label{eq:77}
  S_{\mathcal{I}}S_{\mathcal{I}}^{\dag} = 1, \qquad S_{\mathcal{I}}^{\dag} S_{\mathcal{I}} = P_{\mathcal{I}},
\end{equation}
where $P_{\mathcal{I}}$ is the projection operator on the ideal
$\mathcal{I}$. The matrix $\Delta^{\dag}$ contains the information
about the resolution of the ideal corresponding to the solution of the
ADHM equations. Consider for example the 1-instanton
solution~(\ref{eq:73}). It corresponds to the ideal
$\mathcal{I}_{[[[1]]]}$ of polynomials without constant terms. The
resolution of this ideal is written as the following exact sequence:
\begin{equation}
  \label{eq:108}
  0 \to \mathcal{O} \stackrel{\mu_4}{\to} \mathcal{O}^{\oplus 4} \stackrel{\mu_3}{\to} \mathcal{O}^{\oplus 6}
  \stackrel{\mu_2}{\to} \mathcal{O}^{\oplus 4} \stackrel{\mu_1}{\to} \mathcal{O} \stackrel{p}{\to} \mathcal{I}_{[[[1]]]}
  \to 0,
\end{equation}
where $\mathcal{O} = \mathbb{C}[z_1, z_2, z_3, z_4]$, $p$ is the
projection and the linear operators $\mu_i$ are
\begin{gather}
  \mu_1 = (
  \begin{array}{cccc}
    z_1 & z_2 & z_3 & z_4
  \end{array}
),\\
\mu_2 = \left(
  \begin{array}{cccccc}
    z_2 & z_3 & z_4 & 0 & 0 & 0\\
    -z_1 & 0 & 0 & z_3 & -z_4 & 0\\
     0& -z_1 & 0 & -z_2 & 0 & z_4\\
     0 & 0& -z_1 & 0 & z_2 & -z_3
  \end{array}
\right),\\
\mu_3 = \left(
  \begin{array}{cccc}
    0& 0& z_4 & z_3 \\
    0& -z_4 & 0 & z_2\\
     0& z_3 & -z_2 & 0\\
     z_4 &  0 & 0 & -z_1\\
     z_3 &  0 & -z_1 &0 \\
     z_2 &  -z_1& 0  &0 
  \end{array}
\right),\qquad \qquad 
\mu_4 = \left(
  \begin{array}{c}
    z_1\\
    z_2\\
    z_3\\
    z_4\\
    \end{array}
\right).
\end{gather}
Notice that these operators are very similar to those featuring in
$\Delta^{\dag}$ in Eq.~(\ref{eq:103}). This similarity seems to be a
generic property of any ADHM-like construction.

We are interested in the solutions which are fixed points of the
$\Unitary(1)^3 \subset \SU(4)$ action on $\mathbb{C}^4$. Those correspond to
monomial ideals in the ring $\mathbb{C}[z_1,z_2,z_3,z_4]$ and are
enumerated by solid partitions. We denote the ideal corresponding to a
solid partition $\sigma$ by $\mathcal{I}_{\sigma}$. The fixed point
multi-instanton solutions can be obtained with an ansatz similar to
Eq.~(\ref{eq:73}), but now the function $f(r)$
does not need to be symmetric in $N_a = a_a^{\dag} a_a$ (no summation over $a$), so
that
\begin{equation}
  \label{eq:78}
  Z_a = U_{\sigma} z_a f^{(\sigma)}_a(N_1,N_2, N_3, N_4) U_{\sigma}^{\dag}
\end{equation}
We thus have to determine four functions $f^{(\sigma)}_a(N_1, N_2,
N_3, N_4)$ of four variables. Eqs.~(\ref{eq:76}),~(\ref{eq:70}) imply the following
recurrence relations\footnote{We denote by $f_a(N_b+1)$ the function $f_a(N)$ with $N_b\to N_b+1$.} for $f_a$
\begin{gather}
  \label{eq:79}
  f_a(N_b +1) f_b(N) = f_b(N_a +1) f_a(N),\\
  \sum_{a=1}^4 \left\{ (f_a(N))^2 (N_a + 1) - (f_a(N_a -1))^2 N_a \right\}
  = 4. \label{eq:80}
\end{gather}
Eqs.~(\ref{eq:79}) are ``flatness'' conditions for $f_a$ and can be
solved explicitly. Indeed, one can see that
\begin{equation}
  \label{eq:81}
  f_a(N) =  \frac{h(N)}{h(N_a + 1)}
\end{equation}
solves Eqs.~(\ref{eq:79}) for any function $h(N)$. There remains a
single recurrence relation~(\ref{eq:80}) for $h(r)$, which reads
\begin{equation}
  \label{eq:106}
  \sum_{a=1}^4 \left\{ \frac{h(N)^2}{h(N_a+1)^2} (N_a+1) -
    \frac{h(N_a-1)^2}{h(N)^2} N_a \right\} = 4.
\end{equation}

An example of instanton solution with second Chern class equal to $\frac{k(k+1)(k+2)(k+3)}{24}$ is given by
\begin{equation}
 \label{eq:107}
   Z_a = S_k z_a f^{(k)}(N) S_k^{\dag},
\end{equation}
where $S_k$ now avoids all the states corresponding to monomials of
degree at most $k$ and
\begin{equation}
 \label{eq:109}
 f^{(k)}(N) = \left( 1 - \frac{k(k+1)(k+2)(k+3)}{N(N+1)(N+2)(N+3)} \right)^{\frac{1}{2}}.
\end{equation}
Notice that all four functions $f_a(N)$ are in this case equal to each
other. The solutions~(\ref{eq:109}) correspond to ``pentachoron''
solid partitions (decreasing sequences of tetrahedral plane
partitions), e.g.\ $[[[1]]]$ for $k=1$ or $[[[2,1],[1]],[[1]]]$ for
$k=2$.
\section{Counting solid partitions on orbifolds}\label{4}
\subsection{Orbifolding the quiver}\label{sec:adhm-data-decomposition}

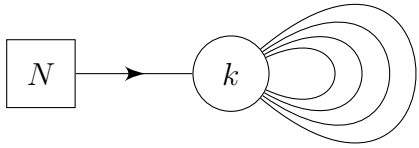
\begin{figure}[ht]
    \centering\vspace{-15mm}
    \begin{tikzpicture}
    \node[FrameNode](F0) at (0,0){$N$};
    \node[GaugeNode](Gk0) at (2.5,0){$k$};
    \draw[->-](F0) to (Gk0);
    \draw[-](Gk0) to[out=-40,in=40,looseness=14] (Gk0);
    \draw[-](Gk0) to[out=-35,in=35,looseness=12] (Gk0);
    \draw[-](Gk0) to[out=-30,in=30,looseness=10] (Gk0);
    \draw[-](Gk0) to[out=-25,in=25,looseness=8] (Gk0);
    \end{tikzpicture}\vspace{-15mm}
    \caption{Local model for the Quot scheme of points.}
    \label{fig:quot-quiver}
\end{figure}

As we showed in section \ref{sec:adhm-construction-8d}, eight-dimensional instanton dynamics is encoded in the representation theory of the quiver depicted in Fig.\ref{fig:quot-quiver}, with relations
\begin{equation}\label{eqn:quiver-relations}
    [B_a,B_b]=0,\qquad 1\le a<b\le 4
\end{equation}
and a stability condition.\footnote{Strictly speaking we showed that the moduli space of $SU(4)$ instantons can be identified with a space of matrices cut by equations \eqref{eqn:quiver-relations} plus an additional real constraint, modulo gauge symmetry. Though it have not rigorously been proved as of this writing, it is believed that the last real condition can be traded for a stability condition, so that the moduli space of instantons may be identified with stable representations of a quiver with relations.}
The moduli space of its stable representations $\MM_{k,N}$ is isomorphic to the quot scheme of points $\Quot_{\mathbb A^4}(\OO^{\oplus N},k)$ or, equivalently, to the moduli space of framed torsion free sheaves on $\mathbb P^4$, as we briefly show in appendix \ref{appendix:framed sheaves}. This isomorphism follows from an application of Beilinson's theorem or, equivalently, from an infinitesimal argument due to \cite{Cazzaniga2020}.

The next natural step is then to study eight-dimensional instanton dynamics on orbifolds of $\mathbb C^4$. Then we let $G$ be a finite subgroup $G\subset\SU(4)$, and study instantons on $\mathbb C^4/G$. 
From the open string theory perspective this amounts to consider twisted representations of the Chan-Paton factors under the discrete group $G$, which manifest in the low energy quiver dynamics of fractional and regular branes \cite{Douglas1996}. 
The mathematical counterpart of the quiver ADHM-like description for orbifold instantons on $\mathbb C^4$ can be obtained as an application of Beilinson's theorem which is outlined in appendix \ref{appendix:orbifold beilinson}.

The useful thing to point out here is that the monad description for the moduli space $\MM_{k,N}^G$ of orbifold instantons, which can be obtained by means of homological algebra, is then given in terms of a sequence of maps between vector spaces. These maps can be easily understood as an equivariant decomposition (in terms of the $G$-action) of the maps and vector spaces arising in the quiver description of $\MM_{k,N}$. If we introduce the action of $G$ on the coordinates $z_\alpha$ in $\mathbb C^4$ by $r_\alpha z_\alpha$, we have a decomposition of the fundamental representation $Q=\rho_{r_1}\oplus\cdots\oplus\rho_{r_4}$, where $\rho_{r_\alpha}$ denotes the irreducible representation of $G$ with weight $r_\alpha$. This decomposition also defines a coloring $\mathbb N^{\oplus 4}\to G$ by
\[
  (n_1,n_2,n_3,n_4)\mapsto\rho_{r_1}^{\otimes n_1}\otimes\rho_{r_2}^{\otimes n_2}\otimes\rho_{r_3}^{\otimes n_3}\otimes\rho_{r_4}^{\otimes n_4}.
\]
Correspondingly we also have decompositions of the vector spaces
\begin{align*}
    &W=\bigoplus_{r}W_r\otimes\rho_r^\vee,
    &V=\bigoplus_{r}V_r\otimes\rho_r^\vee,
\end{align*}
where all of the $W_r$, $V_r$ are finite dimensional vector spaces carrying a trivial $G$-action. The corresponding decomposition of the dimensions $k=\dim_{\mathbb C}V$, $N=\dim_{\mathbb C}W$ is then induced as 
\begin{align*}
    & k=\sum_rk_r=\sum_r\dim_{\mathbb C}V_r,
    & N=\sum_rN_r=\sum_r\dim_{\mathbb C}W_r.
\end{align*}
Here $k_r=\dim_{\mathbb C}V_r$ represents the fractional instanton charge in the $\rho_r^\vee$ representation of $G$, which, from an equivariant localisation point of view, will specify the number of boxes of $r$-th type in a $G$-colored solid partition. On the other hand, the gauge sheaf at infinity transforms in a given representation $\rho$ of $G$, and the $N_r$ dimensions determine the multiplicities of the decomposition of $\rho$ in irreducible representations. If the theory is abelian, \ie $N=1$, only one of the $N_r$ is not zero, and equal to one, while in the case of a non-abelian theory one is given with a plethora of different possibilities. We will restrict our attention to the abelian case for the moment. The decomposition of $V$ then induces a decomposition of the linear maps $B_\alpha\in\Hom_G(V,Q\otimes V)$ as
\[
  B=\bigoplus_r(B_1^r,B_2^r,B_3^r,B_4^r),
\]
so that $B_\alpha^r:V_r\to V_{r+r_\alpha}$. This decomposition immediately gives us the orbifold generalisation of the 8d ADHM equations as
\begin{equation}\label{eqn:orbifold_ADHM}
    B_\beta^{r+r_\alpha}B_\alpha^r=B_\alpha^{r+r_\beta}B_\beta^r.
\end{equation}
In general, even for abelian theories, one should consider different cases (corresponding in the abelian case to which one of the $W_r$ vector spaces has non-vanishing dimension). These different choices correspond to instanton configurations with different asymptotics at infinity. One can however argue along the lines of \cite{Cirafici2011} that moduli spaces corresponding to different asymptotics are isomorphic, so that when computing partition functions we will just consider the distinguished boundary condition $\overline N=(1,0,\dots,0)$.

As in the lower dimensional cases (see \cite{Kronheimer1990,Nakajima2007,Cirafici2011}) all the information about the moduli space of orbifold instantons can be encoded in the datum of a quiver generalising the McKay quiver, which is determined by the representation theory data of the $G$-action. This quiver will moreover encode the decomposition of the usual ADHM data according to the $G$-action and it will have the orbifold ADHM equations as relations. One then starts considering all the irreducible representations $\widehat G$ of $G\subset\SL(4,\mathbb C)$. To each representation in $\widehat G$, including the trivial one, we associate a node in the quiver, while a node $r$ is connected to a node $s$ by a number of arrows $a_{rs}$ which is determined by the decomposition
\[
  Q\otimes\rho_r=\bigoplus_s a_{rs}\rho_s,\qquad a_{rs}=\dim_{\mathbb C}\Hom_G(\rho_s,Q\otimes\rho_r).
\]
To the resulting quiver $\mathsf Q$ we also associate the framed quiver $\mathsf Q^f$, its path algebra $\mathbb C\mathsf Q^f$, and the bounded quiver $(\mathsf Q^f,\mathsf R)$ determined by an ideal $\langle\mathsf R\rangle$ of relations in $\mathbb C\mathsf Q^f$. Its representations form the category $\Reps(\mathsf Q^f,\mathsf R)$, which is equivalent to the $\Amod$ category of left $\mathsf A$-modules for the factor path algebra $\mathsf A=\mathbb C\mathsf Q^f/\langle\mathsf R\rangle$. Moreover, to each vertex $v$ of $\mathsf Q$ one can associate a simple module $\mathsf D_v$, defined to be the representation $V_v\cong\mathbb C$ and $V_w=0$, for $v\neq w$. Projective resolutions of the simple modules $\mathsf D_v$ can be constructed by means of the submodule $\mathsf P_v$ of $\mathsf A$ generated by paths beginning on vertex $v$
\[
\begin{tikzcd}
 \cdots\arrow[r] & \bigoplus_w d^p_{w,v}\mathsf P_w\arrow[r] & \cdots\arrow[r] & \bigoplus_w d^1_{w,v}\mathsf P_w\arrow[r] & \mathsf P_v\arrow[r] & \mathsf D_v\arrow[r] & 0,
\end{tikzcd}
\]
where
\[
d^p_{v,w}=\dim_{\mathbb C}\ext^p_{\mathsf A}(\mathsf D_v,\mathsf D_w).
\]
In the lower dimensional case a special role is played by representations $\Reps(\mathsf Q,\mathsf R)$ of the bounded quiver with dimensions $k_r=1$ and $\dim_{\mathbb C}W=1$, as it turns out they correspond to smooth crepant resolutions of toric singularities. In some cases also the path algebra $\mathsf A$ is a different desingularisation by itself, known as the noncommutative crepant resolution of the toric singularity, which contains the coordinate ring of the singularity as its center. In 
four complex dimensions, however, a crepant resolution of the orbifold singularity is not even granted to exist, though in some simple classes of  examples this is known to be the case \cite{sato2010,satoy2019}. 
Take as an example the case of $\mathbb C^4/\mathbb Z_4$ with the diagonal action $(z_1,z_2,z_3,z_4)\mapsto(\zeta z_1,\zeta z_2,\zeta z_3,\zeta z_4)$, where $\zeta=e^{2\pi i/4}=i$. As $Q=\rho_1\oplus\rho_1\oplus\rho_1\oplus\rho_1$, $r_\alpha=1$ for each $\alpha=1,\dots,4$ and the relevant associated quiver $\mathsf Q$ is then
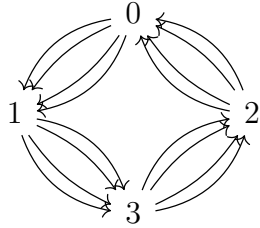
\[
\begin{tikzcd}
 & 0\arrow[dl, bend left = 15]\arrow[dl, bend left = 30]\arrow[dl, bend right = 15]\arrow[dl, bend right = 30] \\
1\arrow[dr, bend left = 15]\arrow[dr, bend left = 30]\arrow[dr, bend right = 15]\arrow[dr, bend right = 30] & & 2\arrow[ul, bend left = 15]\arrow[ul, bend left = 30]\arrow[ul, bend right = 15]\arrow[ul, bend right = 30] \\
 & 3\arrow[ur, bend left = 15]\arrow[ur, bend left = 30]\arrow[ur, bend right = 15]\arrow[ur, bend right = 30]
\end{tikzcd}
\]
and the maps $B_\alpha^r:\mathsf D_r\to\mathsf D_{r+r_\alpha\mod 4}$ are of the form
\begin{align*}
& B_\alpha^0:\mathsf D_0\to\mathsf D_1,\\
& B_\alpha^1:\mathsf D_1\to\mathsf D_2,\\
& B_\alpha^2:\mathsf D_2\to\mathsf D_3,\\
& B_\alpha^3:\mathsf D_3\to\mathsf D_0.\\
\end{align*}
The relevant relations for the unframed quiver are obtained by decomposing accordingly the ADHM equations, thus obtaining
\begin{align*}
    B^1_2B^0_1=B^1_1B^0_2,\qquad 
    B^1_3B^0_1=B^1_1B^0_3,\qquad 
    B^1_4B^0_1=B^1_1B^0_4,\qquad 
    B^1_3B^0_2=B^1_2B^0_3,\\
    B^1_4B^0_2=B^1_2B^0_4,\qquad 
    B^1_4B^0_3=B^1_3B^0_4,\qquad
    B^2_2B^1_1=B^2_1B^1_2,\qquad 
    B^2_3B^1_1=B^2_1B^1_3,\\
    B^2_4B^1_1=B^2_1B^1_4,\qquad 
    B^2_3B^1_2=B^2_2B^1_3,\qquad
    B^2_4B^1_2=B^2_2B^1_4,\qquad 
    B^2_4B^1_3=B^2_3B^1_4,\\
    B^3_2B^2_1=B^3_1B^2_2,\qquad 
    B^3_3B^2_1=B^3_1B^2_3,\qquad 
    B^3_4B^2_1=B^3_1B^2_4,\qquad 
    B^3_3B^2_2=B^3_2B^2_3,\\
    B^3_4B^2_2=B^3_2B^2_4,\qquad 
    B^3_4B^2_3=B^3_3B^2_4,\qquad
    B^0_2B^3_1=B^0_1B^3_2,\qquad 
    B^0_3B^3_1=B^0_1B^3_3,\\ 
    B^0_4B^3_1=B^0_1B^3_4,\qquad 
    B^0_3B^3_2=B^0_2B^3_3,\qquad
    B^0_4B^3_2=B^0_2B^3_4,\qquad 
    B^0_4B^3_3=B^0_3B^3_4.
\end{align*}
The center $\mathsf Z(\mathsf A)$ of the path algebra $\mathsf A$ associated to the bounded quiver is generated as a ring by elements
\[
  \mathsf x_{\alpha\beta\gamma\delta}=B^3_\alpha B^2_\beta B^1_\gamma B^0_\delta,\qquad\alpha\le\beta\le\gamma\le\delta.
\]
As the $G$-action is chosen to be diagonal one can identify the generators $\mathsf x_{\alpha\beta\gamma\delta}$ with the invariant elements in $\mathbb C[z_1,z_2,z_3,z_4]$ by $\mathsf x_{\alpha\beta\gamma\delta}\rightsquigarrow z_\alpha z_\beta z_\gamma z_\delta$, so that
\[
  \spec\mathsf Z(\mathsf A)\cong\mathbb C^4/\mathbb Z_4
\]
 and the factor path algebra $\mathsf A$ is a resolution of the orbifold singularity $\mathbb C^4/\mathbb Z_4$.

\subsection{Orbifold partition function}
The K-theoretic instanton partition function in eight dimensions has been studied in \cite{Nekrasov2017} by means of supersymmetric localisation in terms of the quantum mechanics of a D0-D8 system. In the abelian case the moduli space of BPS vacua is identified with the Hilbert scheme of points of $\mathbb C^4$. In the general case of a proper Calabi-Yau fourfold $X$, the Hilbert scheme of points $X^{[n]}$ is known to carry an obstruction theory, though not perfect. The mathematical definition of the DT-like invariants corresponding to the instanton partition function is made difficult precisely by the latter fact. It is known that they depend on the choice of an orientation of the virtual tangent space and that they need insertions in order to be defined properly. Indeed, if $\mathcal Z\subseteq X^{[n]}\times X$ denotes the universal object, then the virtual tangent space to $X^{[n]}$ can be written as
\[
    T^{\rm vir}_{X^{[n]}} = \RHom_{\pi_{X^{[n]}}}(I_\mathcal Z, I_\mathcal Z)_0[1] = \bold R\pi_{X^{[n]},\ast}\circ\RHom(I_\mathcal Z, I_\mathcal Z)_0[1],
\]
and this obstruction theory is not perfect. However, the machinery put forward by the work of Borisov-Joyce, \cite{Borisov2017}, and more recently by Thomas \cite{Oh:2020rnj}, one can still construct a virtual fundamental class $[X^{[n]}]^{\rm vir}_{o(\mathcal L)}$ depending on the choice of a square root of the isomorphism $Q:\mathcal L\times\mathcal L\to \OO$, where $\mathcal L = \det\RHom_{\pi_{X^{[n]}}}(I_\mathcal Z, I_\mathcal Z)$. As, however, we are interested in the case of a quasi-projective variety, namely $(\mathbb C^4)^{[n]}$, the previous observations don't provide direct access to definitions of relevant invariants. Equivariant localisation (with respect to the action of $\mathbb T=\{(t_1,t_2,t_3,t_4)\in(\mathbb C^\ast)^4:t_1t_2t_3t_4=1\}\subset(\mathbb C^\ast)^4$), however, does provide an easy way out. For a thorough description of this procedure, see \cite{Cao2019a}. For us, let it suffice to say that, given any $\mathbb T$-equivariant line bundle $L$ on $X$, one can define the following K-theoretic invariant, with a slight abuse of notation
\[
\begin{split}
    Z_X^K(L,y)&=\chi\left(X^{[n]},\widehat{\mathscr O}^{\rm vir}\otimes\frac{\bigwedge^\bullet(L^{[n]}\otimes y^{-1})}{\det^{1/2}(L^{[n]}\otimes y^{-1})}\right)\\
    &=\sum_{S\in (X^{[n]})^{\mathbb T}}(-1)^{o(\mathcal L)|_S}e\left(\sqrt{\Ob_{X^{[n]}}|_S}^{\rm fix}\right)\frac{\ch\left(\sqrt{K^{\rm vir}_{X^{[n]}}|S}^{1/2}\right)}{\ch\left(\bigwedge^\bullet\sqrt{N^{\rm vir}|_S}^\vee\right)}\cdot\\
    &\qquad\cdot\frac{\ch\left(\bigwedge^\bullet(L^{[n]}\otimes y^{-1})\right)}{\ch\left(\det^{1/2(L^{[n]}\otimes y^{-1})}\right)}\td\left(\sqrt{T^{\rm vir}_{X^{[n]}}|_S}^{\rm fix}\right),
\end{split}
\]
where $L^{[n]}=\bold R\pi_{X^{[n]},\ast}(\bold R\pi_X^\ast L\otimes\OO_{\mathcal Z})$ and the choice of the square root of the virtual tangent space at a fixed point $S$ induces those of $\Ob_{X^{[n]}}|_S=h^1(T^{\rm vir}_{X^{[n]}})$, $K^{\rm vir}_{X^{[n]}}|_S=\det(T^{\rm vir}_{X^{[n]}}|_S^\vee)$ and $N^{\rm vir}|_S=(T^{\rm vir}_{X^{[n]}}|_S)^{\rm mov}$. Before moving on, let us also notice that the choice that square roots is not unique, so that the invariants at hand are defined only up to a sign. Precisely this definition of $Z_X^K(L,y)$ is what the partition function of the D0-D8 system computes, with a given prescription for the orientation choice.

All these considerations translate into the physical treatment of the problem, where the supersymmetric measure corresponding to the bulk contribution to the Witten index manifest ghost number anomaly, reminiscent of the positive virtual dimension of the underlying moduli space. Then, in the absence of the $\Omega$ deformation, the Witten index is vanishing unless observables matching the ghost number anomaly are inserted. This can be neatly done by adding auxiliary hypermultiplets representing the matter deformation necessary in order to cure the anomaly. A similar story goes for the non-abelian case, which generalises the moduli space to the Quot scheme of points of $\mathbb C^4$, and was studied in \cite{Nekrasov2019}. In general the partition function takes the form
\[
Z^{D8}_N=\sum_{k}Z^{D8}_{N,k}q^k,
\]
where $Z_{N,k}^{D8}$ is computed by the JK integration
\[
Z^{D8}_{N,k}=\JKInt Z^{1-{\rm loop}}_{N,k}\de^{k} u=\frac{1}{k!}\sum_{u_\ast\in\mathfrak M_{\rm sing}}\JKRes_{u_\ast,\zeta}\chi_k\de^k u
\]
and the instanton measure $\chi_k$ is defined by
\begin{displaymath}
\begin{split}
    \chi_k \propto & \prod_{i>j}\frac{\sin^2(u_i-u_j)\prod_{a<b}^3\sin(u_i-u_j-\epsilon_a-\epsilon_b)}{\prod_{a=1}^3\sin(u_i-u_j-\epsilon_a)\sin(u_j-u_i-\epsilon_a)}\prod_{i=1}^k\prod_{\alpha=1}^N\frac{\sin(u_i-m_\alpha)}{\sin(u_i-a_\alpha)}.
\end{split}
\end{displaymath}
It is then known that poles contributing to the JK integration are only those corresponding to $N$-tuples of solid partitions $\overline\pi=(\pi_1,\dots,\pi_N)$, such that $|\overline\pi|=|\pi_1|+\cdots+|\pi_N|=k$. It turns out it is more convenient to work with exponential variables $t_a=e^{2i\epsilon_a}$, $x_i=e^{2iu_i}$, $\nu_{\alpha}=e^{2ia_\alpha}$ and $\mu_{\alpha}=e^{2im_\alpha}$, in which case we have $Z_{N,k}^{D8}=\sum_{|\overline\pi|=k}M_{\bold z}(\overline\pi)$, with
\[
M_{\bold z}(\overline\pi)=\Res_{x=x_{\overline\pi}}\chi_k\prod_i\frac{\de x_i}{x_i}
\]
and
\[
\chi_k = \prod_{i\neq j}\frac{(x_j-x_i)\prod_{a<b}^3(x_j-x_it_at_b)}{\prod_{a=1}^4(x_j-x_it_a)}\prod_{i=1}^k\prod_{\alpha=1}^N\frac{\mu_\alpha-x_i}{\nu_\alpha-x_i}
\]
up to a normalisation constant. It also turns out that a more geometric interpretation for the index computation is available. Indeed, in the rank one case, if $\mathsf Q$ denotes the character of a solid partition and $\sqrt{\mathsf V} = \sum_\mu t^\mu-\sum_\nu t^\nu\in K_0^{\mathbb T}(\pt)$ is the character of the square root of the virtual tangent space to the BPS moduli space at a fixed point, one has
\[
\sqrt{\mathsf V}=\mathsf Q-\overline P_{123}\overline{\mathsf Q}\mathsf Q,
\]
where $P_{123}=(1-t_1)(1-t_2)(1-t_3)$ and the involution acts on the generators of $K_0^{\mathbb T}(\pt)$ as $\overline t_i=t_i^{-1}$. Then
\[
M_{\bold z}(\pi)=(-1)^{h(\pi)}\left[-\sqrt{\widetilde{\mathsf V}}\right],
\]
with
\[
h(\pi)=|\pi|+\#\{(a,d):(a,a,a,d)\in\pi\ {\rm and}\ a\le d\},
\]
$\sqrt{\widetilde{\mathsf V}}=\sqrt{\mathsf V}-y\overline{\mathsf Q}$, while the action of the brackets operator $[\bullet]$ on $\bullet\in K_0^T(\pt)$ is defined as
\[
[\mathsf V] = \frac{\prod_\mu[t^\mu]}{\prod_\nu[t^\nu]}= \frac{\prod_\mu(t^{\mu/2}-t^{-\mu/2})}{\prod_\nu(t^{\nu/2}-t^{-\nu/2})}.
\]

The same procedure might be followed in order to compute partition functions for orbifold instantons. In this case, however, the bosonic field content is the one associated to the morphisms of the quiver specifying the natural crepant resolution of the orbifold singularity, provided it exists. Let us notice here that as $G\subset\SL(4,\mathbb C)$ is contained in the localising torus $\mathbb T$ the locus on which the computation localises can be identified with the $G$-invariant part of the $\mathbb T$-fixed one. Moreover, as the geometrical interpretation is that of an equivariant count of $G$-equivariant zero-dimensional schemes, in order to perform computation one can proceed by simply extracting the $G$-invariant part $\sqrt{\widetilde{\mathsf V}}^G$ of $\sqrt{\widetilde{\mathsf V}}$.

In the case of an orbifold theory the JK residue form gets easily generalised. Let us consider for the sake of simplicity the case of $\mathbb C^2/\mathbb Z_n\times\mathbb C^2$. The bosonic field content is now encoded in the relevant quiver describing the resolution of the orbifold. Let then $\mathsf Q_0$ and $\mathsf Q_1$ be the node set and the edge set of the quiver. We have
\[
    Z^{\rm orb}_{\overline N}=\sum_{\overline k}\underline q^{\overline k}Z^{\rm orb}_{\overline N,\overline k},
\]
with $\underline q^{\overline k}=\prod_{a\in\mathsf Q_0}q_a^{k_a}$ and
\[
  Z^{\rm orb}_{\overline N,\overline k}=\JKInt Z^{1-{\rm loop}}_{\overline N,\overline k}\de^{\overline k}\overline x=\sum_{u_\ast\in\mathfrak M_{\rm sing}}\JKRes_{u_\ast,\zeta}\chi^{\rm orb}_{\overline k}\de^{\overline k}\overline x,
\]
where we denote by $\overline x$ the collection of coordinates associated to the gauge nodes in $\mathsf Q^f$, \ie $\overline x=(x^{(0)}_1,\dots,x^{(0)}_{k_0},\dots)$ and
\[
  \de^{\overline k}\overline x=\prod_{a\in\mathsf Q_0}\prod_{i=1}^{k_a}\frac{\de x_i^{(a)}}{x_i^{(a)}}.
\]
The orbifold instanton measure $\chi^{\rm orb}_{\overline k}$ can be easily read off the edge set $\mathsf Q_1$ of the quiver, and we have:
\[
  \chi^{\rm orb}_{\overline k}\propto\prod_{a\in\mathsf Q_0}\frac{1}{k_a!}Z_{\rm f/af}^{(a)}Z_{\rm adj}^{(a)}Z_{\rm bif}^{(a)},
\]
where $Z_{\rm f/af}^{(a)}$, $Z_{\rm adj}^{(a)}$, $Z_{\rm bif}^{(a)}$ encode the bosonic field content of the theory in the fundamental/antifundamental, adjoint and bifundamental representation of the $a$-th node respectively. In particular we have
\begin{align*}
    & Z_{\rm f/af}^{(a)} = \prod_{i=1}^{k_a}\prod_{\alpha=1}^{N_a}\frac{\mu^{(a)}_\alpha-x^{(a)}_i}{\nu^{(a)}_\alpha-x^{(a)}_i},\\
    & Z_{\rm adj}^{(a)} = \prod_{i\neq j}^{k_a}\frac{(x^{(a)}_j-x^ {(a)}_i)(x^{(a)}_j-x^ {(a)}_it_1t_2)}{(x^{(a)}_j-x^{(a)}_it_3)
    (x^{(a)}_j-x^{(a)}_it_4)
    }\\
    & Z_{\rm bif}^{(a)} = \prod_{i=1}^{k_a}\prod_{j=1}^{k_{a+1}}\frac{(x^{(a)}_i-x^{(a+1)}_jt_1t_3)(x^{(a)}_i-x^{(a-1)}_jt_2t_3)}{(x^{(a)}_i-x^{(a+1)}_jt_1)(x^{(a)}_i-x^{(a-1)}_jt_2)},
\end{align*}
with the node indices being understood to be $a\pmod{n}$.

\begin{remark}
The JK integral formula for the 4-fold orbifold partition function immediately reduces to the integral formula for orbifold counting on 3-folds after the specialisation $\mu\rightsquigarrow t_4$, as is to be expected.
\end{remark}

\subsection{\texorpdfstring{An example: $\mathbb C^2/\mathbb Z_2\times\mathbb C^2$}{An example: C2/Z2 x C2}}
As an example let us consider a local $\mathbb P^1$ realised as $\Tot_{\mathbb P^1}\left(\OO(-2)\oplus\OO^{\oplus 2}\right)$. This can be understood as the canonical crepant resolution of the orbifold singularity $\mathbb C^4/\mathbb Z_2\cong\mathbb C^2/\mathbb Z_2\times\mathbb C^2$, with the $\mathbb Z_2$-action defined as $(z_1,z_2,z_3,z_4)\mapsto(\zeta z_1,\zeta^{-1}z_2,z_3,z_4)$. The ADHM data associated to the orbifolded $\mathbb C^2$ directions can be decomposed according to the $\mathbb Z_2$-action in irreducible $\mathbb Z_2$ representations as we described in section \ref{sec:adhm-data-decomposition}. The relevant quiver then takes the following form
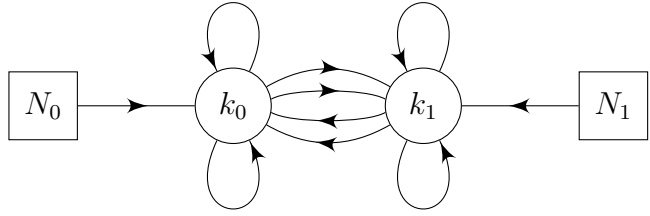
\begin{figure}[H]
\centering
\vspace{-5mm}
\begin{tikzpicture}
\node[FrameNode](F0) at (10,0){$N_0$};
\node[FrameNode](F1) at (17.5,0){$N_1$};
\node[GaugeNode](Gk0) at (2.5+10,0){$k_0$};
\node[GaugeNode](Gk1) at (5+10,0){$k_1$};
\draw[->-](F0) to (Gk0);
\draw[->-](F1) to (Gk1);
\draw[->-](Gk1.10-180) to[bend left, looseness=.5] (Gk0.170-180);
\draw[->-](Gk0.10) to[bend left, looseness=.5] (Gk1.170);
\draw[->-](Gk1.30-180) to[bend left, looseness=1] (Gk0.150-180);
\draw[->-](Gk0.30) to[bend left, looseness=1] (Gk1.150);
\draw[->](Gk0) to[out=65,in=115,looseness=8] (Gk0);
\draw[->](Gk0) to[out=65+180,in=115+180,looseness=8] (Gk0);
\draw[<-](Gk1) to[out=65,in=115,looseness=8] (Gk1);
\draw[<-](Gk1) to[out=65+180,in=115+180,looseness=8] (Gk1);
\end{tikzpicture}
    \caption{Orbifolded $\mathbb C^4/\mathbb Z_2^{(1,-1,0,0)}$ quiver}
    \label{fig:orbifolded quiver_C4/Z2}
\end{figure}
For the sake of simplicity we will restrict to the case $N=1$, which enforces $\overline N=(1,0)$ by the construction in \cite{Kronheimer1990} and the observations in \ref{sec:adhm-data-decomposition}. Supersymmetric localisation can be exploited in order to compute partition functions. As $\mathbb Z_2$ is a subgroup of the localising torus $\mathbb T\cong(\mathbb C^\ast)^3$, the relevant fixed points will simply be identified to be the $\mathbb Z_2$--invariant locus of the $\mathbb T$--fixed one. In particular, as the $\mathbb T$ and the $\mathbb Z_2$-actions commute and as the $\mathbb T$--fixed locus of the theory on $\mathbb C^4$ is into bijective correspondence with solid partitions, if we take the framing to be in the trivial representation of $\mathbb Z_2$ their $\mathbb Z_2$--analogue will be identified with solid partitions $\pi_{\mathbb Z_2}$ decorated by a $\mathbb Z_2$--coloring, which must be compatible with the action of $\mathbb Z_2$ on $\mathbb C^4$. By identifying solid partitions themselves with $\mathbb Z_2$ representations, this coloring is in fact induced by the coloring $(n_1,n_2,n_3,n_4)\mapsto \rho_{r_1}^{\otimes n_1}\otimes\rho_{r_2}^{\otimes n_2}\otimes\rho_{r_3}^{\otimes n_3}\otimes\rho_{r_4}^{\otimes n_4}$ we described in section \ref{sec:adhm-data-decomposition}. One way we can construct these $\mathbb Z_2$--colored solid partitions goes as follows: let $|\pi|=k$ and fix an ordinary partition $\mu=(\mu_1,\mu_2,\dots,\mu_\ell)$, $|\mu|\le k$, colored according to $\mathbb Z_2$--action, then associate to each one of the boxes of the Young diagram of $\mu$ another partition $\lambda_s$, so that $\sum_{s\in\mu}|\lambda_s|=n$, and the partitions $\lambda_s$ must satisfy a nesting relation in either direction of $\mu$. Graphically we have
\[
p\in{\rm fixed\ locus}\hspace{3mm}\longleftrightarrow\hspace{3mm}\ytableausetup{centertableaux,mathmode, boxsize=2em}\begin{ytableau}*(black!30) \lambda_{1,1} & \lambda_{1,2} & \none[\dots]& *(black!30) \lambda_{1,\mu_1} \\\lambda_{1,1} & *(black!30) \lambda_{1,2} & \none[\dots]\\\none[\vdots] & \none[\vdots]& \none[\vdots] \\*(black!30) \lambda_{\ell,1} & \none[\cdots] & \lambda_{\ell,\mu_\ell}\end{ytableau},
\]
with $\lambda_{i,j}\supset\lambda_{i,j+1}$ and $\lambda_{i,j}\supset\lambda_{i+1,j}$, for $(i,j)\in\mu$. The coloring of the resulting solid partition is then induced by a coloring of the Young diagram $\mu$, where each $\lambda_s$ acquires the same color as the underlying box $s\in\mu$. The main difference from the standard instanton counting consists then in the fact that only $\mathbb Z_2$-invariant boxes in $\pi_{\mathbb Z_2}$ are now going to contribute to the computation of the partition functions.

\begin{example}
Consider the case $N=1$, $k=2$ of the cohomological limit of the K-theoretic partition function we discussed in the previous sections. This cohomological limit can be interpreted geometrically as follows: the Chern character provides a natural transformation from the $\mathbb T$-equivariant K-theory to the $\mathbb T$-equivariant Chow group with rational coefficient by $t_i\mapsto e^{s_i}$, with $s_i=c_1^{\mathbb T}(t_i)$. This natural map can be extended to complex coefficients as $t_i^b\mapsto e^{bs_i}$, $b\in\mathbb C$, and it gives a simple linearisation of the K-theoretic brackets $[\bullet]=\bullet^{1/2}-\bullet^{-1/2}$ operator $\ch[t^{b\mu}]=be_{\mathbb T}(t^\mu)+O(b^2)$. This linearisation property can be employed to define a map $Z^{D8}_{\overline N,\overline k}\mapsto Z^{D8,{\rm coh}}_{\overline N,\overline k}$, which can in turn be identified with the rational limit of the trigonometric partition function, from a physical standpoint. The rational partition function is also given an integral representation in terms of JK residues, which also depends on all possible decompositions of $(k,N)$ in $(\overline k,\overline N)$.

Coming back to the particular case of $(k,N)=(2,1)$, one can see that the only possible solid partitions with two boxes are
\begin{equation}
    \pi_1=\ytableausetup{centertableaux,mathmode, boxsize=2em}\begin{ytableau}{[1^1]} & {[1^1]}\end{ytableau}, \qquad
    \pi_2=\ytableausetup{centertableaux,mathmode, boxsize=2em}\begin{ytableau}{[1^1]}\\{[1^1]}\end{ytableau}, \qquad
    \pi_3=\ytableausetup{centertableaux,mathmode, boxsize=2em}\begin{ytableau}{[2^1]}\end{ytableau}, \qquad
    \pi_4=\ytableausetup{centertableaux,mathmode, boxsize=2em}\begin{ytableau}{[1^2]}\end{ytableau}.
\end{equation}
The same solid partitions may also be visualised in the following way:
\begin{equation}
    \pi_1=\vcenter{\hbox{\begin{tikzpicture}[x=(220:.5cm), y=(-40:.5cm), z=(90:0.4cm)]
\foreach \m [count=\y] in {{1,1}}{
  \foreach \n [count=\x] in \m {
  \ifnum \n>0
      \foreach \z in {1,...,\n}{
        \draw [fill=black!30] (\x+1,\y,\z) -- (\x+1,\y+1,\z) -- (\x+1, \y+1, \z-1) -- (\x+1, \y, \z-1) -- cycle;
        \draw [fill=black!40] (\x,\y+1,\z) -- (\x+1,\y+1,\z) -- (\x+1, \y+1, \z-1) -- (\x, \y+1, \z-1) -- cycle;
        \draw [fill=black!10] (\x,\y,\z)   -- (\x+1,\y,\z)   -- (\x+1, \y+1, \z)   -- (\x, \y+1, \z) -- cycle;  
      }
      \node at (\x+0.5, \y+0.5, \n) {\sf{\tiny{\n}}};
 \fi
 }
}    
\end{tikzpicture}}},\qquad
\pi_2=\vcenter{\hbox{\begin{tikzpicture}[x=(220:.5cm), y=(-40:.5cm), z=(90:0.4cm)]
\foreach \m [count=\y] in {{1},{1}}{
  \foreach \n [count=\x] in \m {
  \ifnum \n>0
      \foreach \z in {1,...,\n}{
        \draw [fill=black!30] (\x+1,\y,\z) -- (\x+1,\y+1,\z) -- (\x+1, \y+1, \z-1) -- (\x+1, \y, \z-1) -- cycle;
        \draw [fill=black!40] (\x,\y+1,\z) -- (\x+1,\y+1,\z) -- (\x+1, \y+1, \z-1) -- (\x, \y+1, \z-1) -- cycle;
        \draw [fill=black!10] (\x,\y,\z)   -- (\x+1,\y,\z)   -- (\x+1, \y+1, \z)   -- (\x, \y+1, \z) -- cycle;  
      }
      \node at (\x+0.5, \y+0.5, \n) {\sf{\tiny{\n}}};
 \fi
 }
}    
\end{tikzpicture}}},\qquad
\pi_3=\vcenter{\hbox{\begin{tikzpicture}[x=(220:.5cm), y=(-40:.5cm), z=(90:0.4cm)]
\foreach \m [count=\y] in {{2}}{
  \foreach \n [count=\x] in \m {
  \ifnum \n>0
      \foreach \z in {1,...,\n}{
        \draw [fill=black!30] (\x+1,\y,\z) -- (\x+1,\y+1,\z) -- (\x+1, \y+1, \z-1) -- (\x+1, \y, \z-1) -- cycle;
        \draw [fill=black!40] (\x,\y+1,\z) -- (\x+1,\y+1,\z) -- (\x+1, \y+1, \z-1) -- (\x, \y+1, \z-1) -- cycle;
        \draw [fill=black!10] (\x,\y,\z)   -- (\x+1,\y,\z)   -- (\x+1, \y+1, \z)   -- (\x, \y+1, \z) -- cycle;  
      }
      \node at (\x+0.5, 1, .8) {\sf{\tiny{1}}};
      \node at (\x+0.5, 1, -.3) {\sf{\tiny{1}}};
 \fi
 }
}    
\end{tikzpicture}}},\qquad
\pi_4=\vcenter{\hbox{\begin{tikzpicture}[x=(220:.5cm), y=(-40:.5cm), z=(90:0.4cm)]
\foreach \m [count=\y] in {{1}}{
  \foreach \n [count=\x] in \m {
  \ifnum \n>0
      \foreach \z in {1,...,\n}{
        \draw [fill=black!30] (\x+1,\y,\z) -- (\x+1,\y+1,\z) -- (\x+1, \y+1, \z-1) -- (\x+1, \y, \z-1) -- cycle;
        \draw [fill=black!40] (\x,\y+1,\z) -- (\x+1,\y+1,\z) -- (\x+1, \y+1, \z-1) -- (\x, \y+1, \z-1) -- cycle;
        \draw [fill=black!10] (\x,\y,\z)   -- (\x+1,\y,\z)   -- (\x+1, \y+1, \z)   -- (\x, \y+1, \z) -- cycle;  
      }
      \node at (\x+0.5, \y+0.5, \n) {\sf{\tiny{2}}};
 \fi
 }
}    
\end{tikzpicture}}},
\end{equation}
where the number on the $(i,j,k)$ box of each plane partition denotes the height of a pile of boxes stacked on $(i,j,k)$ along the $\epsilon_4$ direction. The corresponding fixed points labelled with the coloring corresponding to the orbifold action will then be
\begin{equation}
    \pi_1=\ytableausetup{centertableaux,mathmode, boxsize=2em}\begin{ytableau}*(black!30){[1^1]} & {[1^1]}\end{ytableau}, \qquad
    \pi_2=\ytableausetup{centertableaux,mathmode, boxsize=2em}\begin{ytableau}*(black!30){[1^1]}\\{[1^1]}\end{ytableau}, \qquad
    \pi_3=\ytableausetup{centertableaux,mathmode, boxsize=2em}\begin{ytableau}*(black!30){[2^1]}\end{ytableau}, \qquad
    \pi_4=\ytableausetup{centertableaux,mathmode, boxsize=2em}\begin{ytableau}*(black!30){[1^2]}\end{ytableau}.
\end{equation}
The partition function (equivariant under $\mathbb T_0=(\mathbb C^\ast)^3|_{s_1+s_2+s_3+s_4=0}$) for the non orbifolded theory then reads
\begin{equation}\label{eqn:partition-function-example}
    Z_{1,2}^{D8}(\epsilon;q)=\frac{q^2}{2}\left(\frac{m^2(\epsilon_1+\epsilon_2)^2(\epsilon_1+\epsilon_3)^2(\epsilon_2+\epsilon_3)^2}{\epsilon_1^2\epsilon_2^2\epsilon_3^2\epsilon_4^2}+5\frac{m(\epsilon_1+\epsilon_2)(\epsilon_1+\epsilon_3)(\epsilon_2+\epsilon_3)}{\epsilon_1\epsilon_2\epsilon_3\epsilon_4}\right),
\end{equation}
while the contribution from the regular instanton sector to the orbifold partition function will be
\begin{displaymath}
\begin{split}
    Z_{\rm reg}^{D8,{\rm orb}}(\epsilon;q_0,q_1)=& Z_{(1,0),(2,0)}^{D8,{\rm orb}}(\epsilon;q_0,q_1)+Z_{(1,0),(1,1)}^{D8,{\rm orb}}(\epsilon;q_0,q_1)\\
    =&\frac{q_0^2m\epsilon_{12}}{2\epsilon_3\epsilon_{123}(\epsilon_{12}+2\epsilon_3)}\left(\frac{(\epsilon_3-m)(\epsilon_{12}-\epsilon_3)}{\epsilon_3}+\frac{(\epsilon_{123}+m)(2\epsilon_{12}+\epsilon_{3})}{\epsilon_{123}}\right)\\
    &+\frac{q_0q_1m\epsilon_{12}}{2\epsilon_3\epsilon_{123}(\epsilon_2-\epsilon_1)}\left(\frac{(2\epsilon_2+\epsilon_3)(\epsilon_{13}-\epsilon_2)}{\epsilon_2}-\frac{(2\epsilon_1+\epsilon_3)(\epsilon_{23}-\epsilon_1)}{\epsilon_1}\right),\\
\end{split}
\end{displaymath}
with $\epsilon_i=c_1^{\mathbb T}(t_i)$, $\epsilon_{ij}=\epsilon_i+\epsilon_j$ and similarly $\epsilon_{123}=\epsilon_1+\epsilon_2+\epsilon_3$. The previous formula can be either obtained through an integral formula and iterated residues, or by generalising the geometric correspondence of \cite{Nekrasov2019} to the orbifold case
\[
Z_{\rm reg}^{D8,{\rm orb}}(\epsilon;q_0,q_1)=\sum_{|\pi|=2}(-1)^{h(\pi)}e_{\mathbb T}\left[-\left(\sqrt{T_\pi^{\rm vir}}\right)^{\mathbb Z_2}\right]q_0^{|\pi|_0}q_1^{|\pi|_1},
\]
where $\sqrt{T_\pi^{\rm vir}}$ is a square root of the virtual tangent space $T^{\rm vir}_{(\mathbb C^4)^{[k]}}|_\pi$ and $|\pi|_i$ denotes the number of boxes in $\pi$, seen as $\mathbb Z_2$-modules, transforming in the $\rho_i$-th representation of $\mathbb Z_2$.
\end{example}

\vskip 0.5cm
\noindent {\large {\bf Acknowledgments}}
\vskip 0.5cm
We would like to thank 
M.~Cirafici, A.~Gorsky, S.~Mironov, S.~Monavari and Al.~Morozov for interesting discussions.
The research of G.B. is partly supported by the INFN Iniziativa Specifica ST\&FI and by the PRIN project ``Non-perturbative Aspects Of Gauge Theories And Strings''.
The research of N.F.\ and A.T.\ is partly supported by the INFN Iniziativa Specifica GAST.
The work of A.T.\ is partly supported by the PRIN project ``Geometria delle variet\`a algebriche''.
Y.Z. is partly supported by RSF grant 18-71-10073.

\appendix

\section{\texorpdfstring{Review of the $4d$ ADHM construction}{Review of the 4d ADHM construction}}
\label{sec:warm-up:-standard}

\subsection{\texorpdfstring{Self-dual connections in $4d$}{Self-dual connections in 4d}}
\label{sec:self-dual-conn}
In this appendix we recall some basic facts about the standard ADHM construction \cite{Atiyah:1978ri} of the
solutions to the self-duality equation on $\mathbb{R}^4$:
\begin{equation}
  \label{eq:1}
  F_{\mu \nu} = \frac{1}{2} \epsilon_{\mu \nu \lambda \rho} F_{\lambda \rho},
\end{equation}
where $F_{\mu \nu} = \partial_{\mu} A_{\nu} - \partial_{\nu} A_{\mu} +
[A_{\mu}, A_{\nu}]$ is the curvature of a $\Unitary(N)$ connection. The
instanton number is defined as
\begin{equation}
  \label{eq:2}
  k = \frac{1}{16 \pi^2} \int d^4 x\, \epsilon_{\mu \nu \lambda \rho} \tr (F_{\mu \nu} F_{\lambda \rho}) 
\end{equation}
We denote the moduli space of self-dual $\Unitary(N)$ connections of
instanton number $k$ by $\MM_{k,N}$. The (virtual) dimension
of the moduli space of instantons is by definition the number of
independent deformations 
\begin{equation}
  \label{eq:3}
  (\delta^{\alpha}_{\mu} \delta^{\beta}_{\nu} - \frac{1}{2}
  \epsilon^{\alpha \beta \mu \nu})D_{\mu} \delta A_{\nu} = 0,
\end{equation}
minus the number of gauge degrees of freedom:
\begin{equation}
  \label{eq:4}
  \delta A_{\mu} \sim D_{\mu} \phi,
\end{equation}
where $D_{\mu}$ is the covariant derivative in the background
$A_{\mu}$. This difference can be understood as the index of the elliptic 
complex (this requires certain vanishing theorem for the
cohomologies in degrees $0$ and $2$):
\begin{equation}
  \label{eq:13}
\mathcal C:\quad 0 \to \Omega^0(\mathbb{R}^4, \mathfrak{su}(N)) \stackrel{D_A}{\to}
 \Omega^1 (\mathbb{R}^4, \mathfrak{su}(N))
 \stackrel{(1-*)D_A}{\to} \Omega^2_{+} (\mathbb{R}^4,
 \mathfrak{su}(N)) \to 0,
\end{equation}
which by the Atiyah-Singer index theorem is given by
\begin{equation}
  \label{eq:5}
 \dim \MM_{k,N} = \mathrm{ind}\, \mathcal{C} = 4 N k.
\end{equation}

\subsection{ADHM data and equations}
\label{sec:adhm-data-equations}
In the ADHM construction we consider a complex $(2k+N) \times N$
matrix $U$ satisfying the equation
\begin{equation}
  \label{eq:6}
  \Delta^{\dag}(x) U(x) = 0,
\end{equation}
where $\Delta(x)$ is a $(2k +N)\times 2k$ complex matrix
\begin{equation}
  \label{eq:8}
  \Delta(x) = \left(
    \begin{array}{cc}
      B_2^{\dag} - z_2^{*} & -B_1 + z_1 \\
      B_1^{\dag} - z_1^{*} & B_2 - z_2\\
      I^{\dag} & J 
    \end{array}
\right),
\end{equation}
and $z_1 = x_2 + ix_1$, $z_2 = x_4 + i x_3$. The auxiliary matrix
$\Delta(x)$ is required to satisfy the moment map equation
\begin{equation}
  \label{eq:9}
  \Delta^{\dag}(x) \Delta(x) = 1_{2\times 2} \otimes f^{-1}_{k\times k}(x),
\end{equation}
where $f(x)$ is a $k\times k$ invertible matrix. Eq.~\eqref{eq:9} leads to the
well-known conditions on $B_{1,2}$, $I$ and $J$:
\begin{equation}
  \label{eq:10}
  \mu_{\mathbb{C}} = [B_1, B_2] + IJ = 0,\qquad \mu_{\mathbb{R}} = [B_1, B_1^{\dag}] + [B_2, B_2^{\dag}] - I
  I^{\dag} + J^{\dag} J =0.
\end{equation}
One has to normalize $U$ so that
\begin{equation}
  \label{eq:7}
  U^{\dag} U = 1_{N \times N}.
\end{equation}
In other words, the matrix $U$ defines a basis of $N$ orthonormal
vectors inside $\mathbb{C}^{2k+N}$. The columns of the matrix
$\Delta(x)$ form a set of $k$ linearly independent vectors inside
$\mathbb{C}^{2k+N}$ (the linear independence is guaranteed by
Eq.~\eqref{eq:9}). Eq.~\eqref{eq:6} means that the basis defined by
$U$ is orthogonal to the set of $2k$ vectors inside
$\mathbb{C}^{2k+N}$ defined by $\Delta(x)$. Together the columns of
$U$ and $\Delta$ form a complete basis in $\mathbb{C}^{2k+N}$, and therefore:
\begin{equation}
  \label{eq:11}
  1_{(2k+N) \times (2k+N)} - U(x) U^{\dag}(x) = \Delta(x) (1_{2\times 2} \otimes
  f_{k \times k}(x)) \Delta^{\dag}(x).
\end{equation}

The self-dual connection $A_{\mu}(x)$, corresponding to the matrix
$U(x)$ is given by
\begin{equation}
  \label{eq:12}
  A_{\mu}(x) = U^{\dag} \partial_{\mu} U(x).
\end{equation}
It is easy to show that the connection~\eqref{eq:12} is indeed
self-dual:
\begin{multline}
  \label{eq:14}
  F_{\mu\nu} = \partial_{\mu} U^{\dag} \partial_{\nu} U(x)
  - \partial_{\nu} U^{\dag} \partial_{\mu} U(x) + [
  U^{\dag} \partial_{\mu} U(x), U^{\dag} \partial_{\nu} U(x)] =\\
  = (\partial_{[\mu} U^{\dag}) (1 - U U^{\dag}) (\partial_{\nu]}
  U^{\dag}) = U^{\dag} (\partial_{[\mu} \Delta(x)) (1_{2\times 2}
  \otimes f_{k \times k}(x))
  (\partial_{\nu]} \Delta^{\dag}(x)) U =\\
  = U^{\dag} (\sigma_{\mu \nu}
  \otimes  f_{k \times k}(x)) U,
\end{multline}
where
\begin{equation}
  \label{eq:15}
  \sigma_{\mu \nu} = \sigma_{\mu} \bar{\sigma}_{\nu} - \sigma_{\nu}
  \bar{\sigma}_{\mu} = i \eta^a_{\mu \nu} \sigma_a,
\end{equation}
and $\sigma_{\mu} = (1, i \vec{\sigma})$, $\bar{\sigma}_{\mu}=
\sigma_{\mu}^{\dag} = (1, - i \vec{\sigma})$, $\eta^a_{\mu\nu}$ is the
't Hooft symbol. The matrices $(\sigma_{\mu\nu})_{\alpha}{}^{\beta}$
when seen as matrices with indices $\mu\nu$ are known to be
self-dual. In fact they give an intertwining operator between the
three-dimensional representation $\Lambda^2_{+}\mathbb{R}^4$ of
$\SO(4)$ and the adjoint representation of $SU(2)_{+}$, which is part
of $\Spin(4) = \SU(2)_{+} \times \SU(2)_{-}$, so that
\begin{equation}
  \label{eq:16}
  \Lambda^2_{\pm}\mathbb{R}^4 = \mathrm{adj}_{\SU(2)_{\pm}}.
\end{equation}
The moduli space $\MM_{k,N}$ of solutions to the ADHM
equations is the space of matrices $(B_1, B_2, I, J)$ obeying
$\mu_{\mathbb{C}}=0$ and $\mu_{\mathbb{R}} = 0$ quotiented by the
action of $\Unitary(k)$ group (indeed, constant $\Unitary(k)$ transformations don't
change the connection~\eqref{eq:12}). The resulting dimension of the
moduli space is
\begin{equation}
  \label{eq:17}
  \dim_{\mathbb{R}} \MM_{k,N} = \underbrace{2 k^2}_{B_1} +
  \underbrace{2 k^2}_{B_2} + \underbrace{2 N k}_I + \underbrace{2 N k}_J -
  \underbrace{2 k^2}_{\mu_{\mathbb{C}}} -   \underbrace{\mu_{\mathbb{R}}}_{k^2} -   \underbrace{k^2}_{/U(k)} = 4 Nk,
\end{equation}
is equal to the dimension of the tangent~\eqref{eq:5}, as it should.

\subsection{Spinor formalism}
\label{sec:spinors}
Let us see how the ADHM equations can be formulated using spinors. Let
$\gamma^{\mu}$, $\mu = 0, \ldots, 4 $ be four-dimensional
gamma-matrices generating the algebra $\Cliff(4)$. Then we have
\begin{equation}
  \label{eq:18}
  \gamma^{\mu}\gamma^{\nu}\gamma^{\lambda}\gamma^{\rho} =
  \epsilon^{\mu \nu \lambda \rho} \gamma_5,
\end{equation}
where $\gamma_5$ is the chirality operator. Let $\psi_{\pm}$ be a
positive (resp.\ negative) chirality spinor of $\Spin(4)$
\begin{equation}
  \label{eq:21}
  \gamma_5 \psi_{\pm} = \pm \psi_{\pm},
\end{equation}
normalized so that $\psi^{\dag}_{\pm} \psi_{\pm} = 1$. We can then
trivially write that
\begin{equation}
  \label{eq:19}
  \psi^{\dag}_{\pm}
  \gamma^{\mu}\gamma^{\nu}\gamma^{\lambda}\gamma^{\rho} \psi_{\pm} =
  \pm  \epsilon^{\mu \nu \lambda \rho}.
\end{equation}
Notice that this identity is independent of the concrete value of
$\psi_{\pm}$. We can thus write the self-duality condition for $F_{\mu
\nu}$ as
\begin{equation}
  \label{eq:22}
  F_{\mu \nu} = \frac{1}{2}   \psi^{\dag}_{+}
  \gamma^{\mu}\gamma^{\nu}\gamma^{\lambda}\gamma^{\rho} \psi_{+}
  F_{\lambda \rho}.
\end{equation}
Alternatively, one can notice that
\begin{equation}
  \label{eq:23}
  \gamma^{\mu\nu} = \frac{1}{2} [\gamma^{\mu} \gamma^{\nu}] = \left(
    \begin{array}{cc}
      \sigma_{\mu \nu} & 0\\
      0 & \bar{\sigma}_{\mu \nu}
    \end{array}
\right) = \left(
    \begin{array}{cc}
      \eta^a_{\mu \nu} i \sigma^a & 0\\
      0 & \bar{\eta}^a_{\mu \nu} i \sigma^a
    \end{array}
\right),
\end{equation}
where $\bar{\sigma}_{\mu \nu} = \bar{\sigma}_{[\mu} \sigma_{\nu]}$, so
that the projection on the self-dual part of $F_{\mu \nu}$ can be
written as
\begin{equation}
  \label{eq:24}
  \gamma^{\lambda}\gamma^{\rho} \psi_{-} 
  F_{\lambda \rho}= 0, \qquad \text{or} \qquad \bar{\sigma}_{\lambda
    \rho} F_{\lambda \rho} = 0. 
\end{equation}
Eq.~(\ref{eq:24}) contains three independent equations on the
components of $A_{\mu}(x)$. The forth equation is provided by the
gauge fixing condition, e.g.\ $\partial_{\mu} A_{\mu} =0$, so that the
total number of components is equal to the total number of (first
order) equations. After imposing the equations no \emph{functional}
degrees of freedom in $A_{\mu}(x)$ remain, and the moduli space of
solutions $\MM_{k,N}$ is finite-dimensional.

\subsection{Hyperk\"ahler reduction and complex structure(s)}
\label{sec:inst-compl-struct}
We have considered self-duality equations in $\mathbb{R}^4$ without
any reference to a specific choice of the complex structure. However,
we can incorporate complex structure in our construction.
$\mathbb{R}^4$ is a hypek\"ahler manifold with three basis complex
structures $I$, $J$ and $K$\footnote{There is in fact a $S^2$ worth of
  complex structures. Indeed, $aI + bJ + c K$ is a complex structure
  as long as $a^2 + b^2 + c^2 = 1$.} satisfying the usual quaternionic
relations. Correspondingly, there are three K\"ahler forms $\omega^I$,
$\omega^J$ and $\omega^K$.

These forms can be used to give the structure of the
  hyperk\"ahler manifold to the space $\mathcal{A}$ of $\Unitary(N)$
  connections on $\mathbb{R}^4$. Indeed, the metric on the space of
  connections is induced from the flat metric on $\mathbb{R}^4$
  \begin{equation}
    \label{eq:43}
    ds^2[\delta_1 A_{\mu}(x), \delta_2 A_{\mu}(y)] =
    \int_{\mathbb{R}^4}   \tr \delta_1 A_{\mu}(x) \delta_2 A_{\mu}(x),
  \end{equation}
  and we can write the symplectic forms on $\mathcal{A}$ as follows
\begin{equation}
  \label{eq:35}
  \Omega^a[\delta_1 A_{\mu}(x), \delta_2 A_{\mu}(y)] =
  \int_{\mathbb{R}^4}  \omega^a \wedge \tr \delta_1 A(x) \wedge  \delta_2 A(x),
\end{equation}
where $a = I,J,K$. All three $\omega^a$ are actually \emph{self-dual}
two-forms. One can then verify that the action of $\mathcal{G}$ on
$\mathcal{A}$ is Hamiltonian for all three symplectic forms
$\Omega^a$, and the corresponding three moment maps are
\begin{equation}
  \label{eq:45}
  \mu^a[\phi(x)] = \int_{\mathbb{R}^4} \omega^a \wedge \tr (\phi(x)  F(x)),
\end{equation}
where $F = dA + [A,A]$ is the field strength. Requiring moment maps to
vanish we get precisely the self-duality equations:
\begin{equation}
  \label{eq:44}
  \omega^a \wedge F = 0.
\end{equation}
Indeed, Eq.~(\ref{eq:44}) mean that the \emph{anti-self-dual} part $F$
vanishes. The space of self-dual connections can be thought of as the
hyperk\"ahler reduction of $\mathcal{A}$ by the group $\mathcal{G}$ of
all gauge transformations. If we introduce $\mathcal{A}_k$ as the
space of connections with instanton number $k$ then we can write
\begin{equation}
  \label{eq:36}
  \MM_{k,N} = \mathcal{A}_k /\!/\!/ \mathcal{G}.
\end{equation}

Let us choose a complex structure $I$, such that $z_1$ and $z_2$
defined as in Eq.~\eqref{eq:8} are the holomorphic coordinates. Only a
subgroup $\Unitary(2) \subset \SO(4)$ of rotations preserves this choice. The
choice of the complex structure $I$ singles out one of the K\"ahler
forms $\omega^I$, which can be written as
\begin{equation}
  \label{eq:33}
  \omega^I = \omega_{\mathbb{R}} = dz^1 \wedge d\bar{z}^1 + dz^2 \wedge d\bar{z}^2.
\end{equation}
The other two forms $\omega^J$ and $\omega^K$ can be recast into the
holomorphic symplectic form (and its conjugate)
\begin{equation}
  \label{eq:34}
  \omega^J + i \omega^K = \omega_{\mathbb{C}} = dz^1 \wedge dz^2.
\end{equation}
The subgroup of $\SO(4)$, which preserves both the complex structure
$I$ and $\omega_{\mathbb{C}}$ is $\SU(2) \subset \Unitary(2)$ (the remaining
$\Unitary(1) \subset \Unitary(2)$ rotates the phase of $\omega_{\mathbb{C}}$). The
field strength $F_{\mu \nu}$ breaks into the following irreducible
representations of $\SU(2)$ (we also list the number of components):
\begin{equation}
  \label{eq:37}
  F_{\mu \nu} =
  \begin{cases}
   F^{(2,0)}_{z_1 z_2} = F_{\mu \nu} (\omega^{-1}_{\mathbb{C}})^{\mu\nu}, & \dim = 1,\\
   F^{(0,2)}_{\bar{z}_1 \bar{z}_2} = F_{\mu \nu} (\bar{\omega}^{-1}_{\mathbb{C}})^{\mu\nu}, & \dim = 1,\\
    F^{(1,1)}_{\omega} = F_{\mu \nu} (\omega^{-1}_{\mathbb{R}})^{\mu\nu}, & \dim = 1,\\
    (F^{(1,1)}_0)_{z_i \bar{z}_j} = F^{(1,1)}_{z_i \bar{z}_j} - \frac{1}{2} F^{(1,1)}_{z_k \bar{z}_l}
    (\omega^{-1}_{\mathbb{R}})^{z_k \bar{z}_l}
    (\omega_{\mathbb{R}})_{z_i \bar{z}_j}, & \dim = 3.
  \end{cases}
\end{equation}
The first three one-dimensional pieces turn out to be self-dual (they
are projections on $\omega^a$), while the three-dimensional
representation is anti-self-dual. Having these identifications it is
easy to see that the moment map equation~\eqref{eq:9} is just the
requirement that the $\Delta^{\dag}\Delta$ lies in the self-dual part
$\mathbf{1}$ of the tensor product of $\SU(2)$ representations
$\bar{\mathbf{2}} \otimes \mathbf{2} = \mathbf{3} \oplus \mathbf{1}$.

\subsection{Non-commutative deformation.}
\label{sec:non-comm-deform}
The instanton moduli space $\MM_{k,N}$ contains
singularities, which correspond to instantons of zero size. A clever
way to regularize the moduli space is to consider instantons on the
non-commutative spacetime $\mathbb{R}^4$ with coordinates satisfying:
\begin{equation}
  \label{eq:54}
  [x^{\mu}, x^{\nu}] = i \theta^{\mu \nu},
\end{equation}
where $\theta^{\mu \nu}$ is a constant 2-form. We further assume that
$\theta^{\mu \nu}$ is anti-self-dual, then with an $\SO(4)$ rotation it
can be aligned with one of the three basis K\"ahler structures, e.g.\
\begin{equation}
  \label{eq:55}
 \theta^{\mu \nu} = \frac{\zeta}{4} \bar{\eta}_{\mu \nu}^3,
\end{equation}
where $\zeta$ is a real non-commutativity parameter which we take to
be positive.

As shown in \cite{Nekrasov:1998ss}, the ADHM construction for the
non-commutative case requires only a minor update. The
expression~\eqref{eq:8} for $\Delta(x)$ is still valid and the
conditions~\eqref{eq:9} still holds. However, since the coordinates no
longer commute,
one of the moment maps is modified:
\begin{equation}
  \label{eq:56}
  \mu_{\mathbb{R}} = [B_1^{\dag}, B_1] + [B_2^{\dag}, B_2] + II^{\dag}
  - J^{\dag} J = \zeta 1_{k \times k}. 
\end{equation}
Notice that after the non-commutative deformation, the instanton
solutions also appear in the $\Unitary(1)$ gauge theory.

In string theory language, the self-dual connections we are studying
correspond to bound states of D0 and D4 branes (or, more generally
D$p$ and D$(p+4)$ branes). Non-commutativity arises if we turn on the
nonzero $B$-field background along the D4 brane.

\subsection{Matrix form of the non-commutative self-duality equations.}
\label{sec:matrix-form-non}

In the noncommutative setting it will be convenient for us to recast
the self-duality equations in matrix form. To this end we introduce
the operator analogues of covariant derivatives
\begin{equation}
  \label{eq:82}
  X^{\mu} = x^{\mu} + i\theta^{\mu\nu}A_{\nu},
\end{equation}
The commutator of covariant derivatives gives the field strength:
\begin{equation}
  \label{eq:83}
  [X^{\mu}, X^{\nu}] = i \theta^{\mu \nu} +  \theta^{\mu \lambda}
  \theta^{\nu\rho}F_{\lambda \rho},
\end{equation}
where the second term in the r.h.s.\ is due to Eq.~(\ref{eq:54}).

The self-duality equations can be rewritten as an equation for the
operators $X^{\mu}$
\begin{equation}
  \label{eq:84}
  [X^{\mu}, X^{\nu}] - i \theta^{\mu \nu} = \frac{1}{2} \epsilon_{\mu \nu
    \lambda \rho} ( [X^{\lambda}, X^{\rho}] - i \theta^{\lambda \rho}) .
\end{equation}
Let us choose a K\"ahler structure on $\mathbb{R}^4$ proportional to
the $B$-field. In the complex structure compatible with $B$ the noncommutativity reads
\begin{equation}
  \label{eq:86}
  [z_a, \bar{z}_b] = - \frac{\zeta}{2} \delta_{ab}, \qquad [z_a, z_b]
  = [\bar{z}_a, \bar{z}_b] = 0.
\end{equation}
We introduce complex covariant derivatives $Z_a$ corresponding to
complex coordinates $z_a$. Eqs.~(\ref{eq:84}) are then written as one
real and one complex equation for $Z_a$ operators:
\begin{gather}
  \label{eq:85}
  [Z_1, Z_2] = 0, \qquad [Z_1^{\dag}, Z_2^{\dag}] = 0,\\
  [Z_1^{\dag}, Z_1] + [Z_2^{\dag}, Z_2] = \zeta.\label{eq:116}
\end{gather}
The commutation relations~(\ref{eq:86}) are of course nothing but a
pair of standard Heisenberg algebras of creation and annihilation
operators, so that
\begin{equation}
  \label{eq:87}
  z_a =  \sqrt{\frac{2}{\zeta}} a_a^{\dag}, \qquad \bar{z}_a =  \sqrt{\frac{2}{\zeta}} a_a,
\end{equation}
and
\begin{equation}
  \label{eq:88}
  [a_a, a_b^{\dag}] = \delta_{ab}, \qquad [a_a, a_b] = [a_a^{\dag},
  a_b^{\dag}] = 0.
\end{equation}
The operators $a_a$, $a_a^{\dag}$ act on the Hilbert space
$\mathcal{H}$, which is spanned by the eigenstates $|n,m\rangle$ of
the number operators:
\begin{equation}
  \label{eq:92}
  N_a = a_a^{\dag} a_a, \qquad N_1 | n,m\rangle = n | n,m\rangle,
  \qquad N_2 | n,m\rangle = m | n,m\rangle.
\end{equation}
$\mathcal{H}$ can be identified with the ring of polynomials in
$a^{\dag}_a$ (or in $z_a$) acting on the vacuum $|0,0\rangle$.

The simplest solution to Eqs.~(\ref{eq:85}) is the vacuum solution
\begin{equation}
  \label{eq:89}
  Z_a = z_a,
\end{equation}
which corresponds to zero $A_{\mu}$ and vanishing instanton charge.

Another example of a solution is the non-commutative $U(1)$ instanton
sitting at the origin of $\mathbb{C}^2$:
\begin{equation}
  \label{eq:90}
  Z_a = \sqrt{\frac{2}{\zeta}} S_{[1]} a_a^{\dag}
  \sqrt{\frac{N(N+3)}{(N+1)(N+2)}} S^{\dag} = 
  \sqrt{\frac{2}{\zeta}}S\sqrt{\frac{N+2}{N}} a_a^{\dag}
  \sqrt{\frac{N}{N+2}} S_{[1]}^{\dag},
\end{equation}
where $N = a_1^{\dag} a_1 + a_2^{\dag} a_2$ and $S_{[1]}^{\dag}$ acts
on $\mathcal{H}$ by relabelling the (infinite number of) basis
vectors, so that the state $|0,0\rangle$ does not belong to its image. An example of such a transformation is
\begin{equation}
  \label{eq:91}
  S_{[1]}^{\dag} |n,m\rangle =
  \begin{cases}
    |n,m\rangle & m \neq 0,\\
    |n+1,m\rangle & m = 0.
  \end{cases}
\end{equation}
The solution~(\ref{eq:90}) is non-singular (all of its matrix elements
are well-defined) and invariant under $\Unitary(2)$ rotations of the
space-time $\mathbb{C}^2$ (the operator $S_{[1]}$ is invariant only up
to a unitary transformation, which can be viewed as a gauge
transformation). We give some more examples of instanton solutions
invariant under $\Unitary(1)^2 \subset \Unitary(2)$ in
~\ref{sec:fixed-point-inst}.

\subsection{\texorpdfstring{$\Unitary(1)$ instantons and ideals}{U(1) instantons and ideals}}
\label{sec:u1-inst-ideals}
Let us recall the correspondence between $\Unitary(1)$ noncommutative
instantons and ideals in the ring of polynomials $\mathbb{C}[z_1,
z_2]$. The matrix $U$ for gauge group $\Unitary(1)$ is a $(2k+1)$-dimensional
column vector, which we write as
\begin{equation}
  \label{eq:93}
  U(z) = \left(
    \begin{array}{c}
      \psi_1(z)\\
      \psi_2(z)\\
      \xi(z)
    \end{array}
\right),
\end{equation}
where $\psi_{1,2}(z)$ are two $k$-dimensional column vectors of
polynomials and $\xi(z)$ is a polynomial. Eq.~(\ref{eq:6}) for $U$ is
then equivalent to two equations
\begin{align}
  \label{eq:94}
  I \xi(z) &= (z_2 - B_2) \psi_1(z) + (z_1 - B_1) \psi_2(z),\\
  0 &= (B_1^{\dag} - \bar{z}_1) \psi_1(z) + (\bar{z}_2 - B_2^{\dag})
  \psi_2(z). \label{eq:95}
\end{align}

Eq.~(\ref{eq:94}) implies (see e.g.\cite{Furuuchi:1999kv} for details) that
$\xi(z)$ belongs to an ideal $\mathcal{I} \subset \mathbb{C}[z_1,
z_2]$ of polynomials which vanish when one subsitutes in them matrices
$B_{1,2}$ instead of the variables $z_{1,2}$. Indeed, acting with
Eq.~(\ref{eq:94}) on the states $|n,m\rangle \in \mathcal{H}$ we get
\begin{equation}
 \label{eq:59}
 \xi(z)| n,m \rangle I  = (z_2 - B_2) \psi_1(z)| n,m \rangle + (z_1 -
 B_1) \psi_2(z)| n,m \rangle.
\end{equation}
After the substitution $z_{1,2} \to B_{1,2}$ we get
\begin{equation}
 \label{eq:62}
 \xi(B_1, B_2)| n,m \rangle I = 0.
\end{equation}
Acting with $B_1^k B_2^l$ on Eq.~(\ref{eq:62}) and using the
commutativity of $B_1$ and $B_2$ we see that $\xi(B_1, B_2)| n,m
\rangle$ is zero on the whole $\mathbb{C}^k$, so it is a zero matrix
and thus $\xi(B_1, B_2) = 0$ as a matrix equation. This determines the
ideal completely. Vice versa, each element $\xi(z)$ of $\mathcal{I}$
by definition can be written as a linear combination from
Eq.~(\ref{eq:94}), but not uniquely: one can shift
\begin{align}
  \label{eq:96}
  \psi_1(z) &\to \psi_1(z) + (z_1 - B_1) \chi(z),\\
  \psi_2(z) &\to \psi_2(z) + (z_2 - B_2) \chi(z),\label{eq:97}
\end{align}
for any column-vector $\chi(z)$. To fix a unique decomposition we have
to gauge fix the symmetry~(\ref{eq:96}),~(\ref{eq:97}), which is done
by requiring Eq.~(\ref{eq:95}) (note that $\bar{z}_a$ act as the
derivatives in $z_a$).

For the single $U(1)$ instanton solution described in the previous
section we have $B_1 = B_2 = 0$, $I = \sqrt{\zeta}$ and
\begin{equation}
  \label{eq:98}
  U(z) = \left(
    \begin{array}{c}
      \bar{z}_2\\
      \bar{z}_1\\
      \frac{1}{\sqrt{\zeta}}(z_1 \bar{z}_1 + z_2 \bar{z}_2)
    \end{array}
\right) \lambda(z).
\end{equation}
For any $\lambda(z)$ the vector~(\ref{eq:98})
solves~(\ref{eq:6}). However to satisfy the normalization
condition~(\ref{eq:7}), we have to put the following normalization
factor instead of $\lambda(z)$:
\begin{equation}
  \label{eq:99}
  \mathcal{N} = \frac{1}{\sqrt{(z_1 \bar{z}_1 + z_2 \bar{z}_2) (z_1 \bar{z}_1
    + z_2 \bar{z}_2 + \zeta)}} S^{\dag}.
\end{equation}
The solution~(\ref{eq:90}) is obtained from the normalized $U(z)$ as
\begin{equation}
  \label{eq:100}
  Z_a = U^{\dag} z_a U, \qquad Z_a^{\dag} = U^{\dag} \bar{z}_a U.
\end{equation}

\subsection{\texorpdfstring{$4d$ fixed point instantons}{4d fixed point instantons}}
\label{sec:fixed-point-inst}
We give here some examples of noncommutative instanton
solutions on $\mathbb{R}^4$ which are equivariant under the $\Unitary(1)^2$ subgroup of
$\SO(4)$. The ideals in $\mathbb{C}[z_1,z_2]$ corresponding to these
solutions are monomial ideals enumerated by Young diagrams.

All the solutions have the form
\begin{equation}
  \label{eq:110}
  Z_a = \sqrt{\frac{\zeta}{2}}S_Y h_Y(N_1, N_2)^{-1} a^{\dag}_a h_Y(N_1, N_2) S_Y^{\dag},
\end{equation}
where $S_Y$ is the partial isometry associated with the monomial ideal
labelled by the Young diagram $Y$, $N_a = a_a^{\dag} a_a$ and
\begin{equation}
  \label{eq:111}
  h_Y(N_1, N_2) = \sqrt{\frac{g_Y(N_1, N_2)}{g_Y(N_1+1, N_2+1)}}.
\end{equation}
Let us display the functions $g_Y(N_1, N_2)$ for some elementary Young diagrams 
\begin{align}
  \label{eq:114}
  g_{[1]} &= N,\\
  g_{[2]} &= N^2 - N_1 + N_2,\\
  g_{[1,1]} &= N^2 + N_1 - N_2 ,\\
  g_{[3]} &= N (N^2 - 3(N_1 - N_2) + 2,\\
  g_{[1,1,1]} &= N (N^2 + 3(N_1 - N_2) + 2,\\
    g_{[2,1]} &= (N-1) N (N+1),\\
    g_{[3,2,1]} &= (N-2) (N-1) N^2 (N+1)(N+2).
    \label{eq:117}
\end{align}
where $N = N_1 + N_2$. The first solution~(\ref{eq:114}) is the
one-instanton solution discussed in
sec.~\ref{sec:matrix-form-non}. One can check that for any function
$g_Y$ given above the matrix equations~(\ref{eq:85}),~(\ref{eq:116})
are indeed satisfied. In fact Eq.~(\ref{eq:85}) is satisfied
automatically by the ansatz~(\ref{eq:110}), and it is only necessary
to check a single recurrence relation for the function $h(N_1, N_2)$:
\begin{equation}
  \label{eq:115}
    \sum_{a=1}^2 \left\{ \frac{h(N)^2}{h(N_a+1)^2} (N_a+1) -
    \frac{h(N_a-1)^2}{h(N)^2} N_a \right\} = 2,
\end{equation}
which turns out to be true. Of the solutions given above
only~(\ref{eq:114}) and (\ref{eq:117}) are invariant under the full
$SU(2)$ rotation symmetry. There exists an infinite family of such
fully symmetric solutions corresponding to triangular Young diagrams
with
\begin{equation}
  \label{eq:118}
  h_{[k,k-1,\ldots, 1]}(N_1, N_2) = \left( 1 - \frac{k(k+1)}{N(N+1)}
  \right)^{\frac{1}{2}} = \left( \frac{(N-k)(N+k+1)}{N(N+1)}
  \right)^{\frac{1}{2}}.
\end{equation}

Notice that $g_Y(i,j)$ vanishes if the box $(i,j)$ lies on the border
of the Young diagram $Y$. This guarantees that the action of $Z_a$ and
$Z_a^{\dag}$ from Eq.~(\ref{eq:110}) is well-defined.

\section{8d instantons and sheaf cohomology}\label{b}
\subsection{Moduli spaces of 8d instantons via Beilinson's Theorem}\label{appendix:framed sheaves}
Here we will briefly study moduli spaces of framed torsion-free sheaves on $\mathbb P^4$ and their relation to spaces of $\SU(4)$ instantons (and their generalisations) on $\mathbb C^4$. Let us first notice that, in general, if $\EE$ is a torsion-free sheaf of rank $N$ on $\mathbb P^4$ with $\ch(\EE)=(N,0,0,0,-k)$, framed along a divisor $D$, there exists a natural sequence of sheaves (for the proof of this result, see \cite{Cazzaniga2020})
\[
\begin{tikzcd}[column sep=small]
    0\arrow[r] & \EE\arrow[r, hook] & \OO^{\oplus N}_{\mathbb P^d}\arrow[r, two heads] & \QQ\arrow[r] & 0,
\end{tikzcd}
\]
and $\QQ$ has finite support in $\mathbb A^4\cong\mathbb P^4/D$. Consider then the moduli space
\begin{equation}\label{enq:moduli-space-sheaves-definition}
\left.\left\{
\begin{aligned}
&\EE\in\Coh(\mathbb P^4),\ \EE\ {\rm torsion-free},\ \rk(\EE)=N,\\
& \begin{tikzcd}[column sep=small]
\EE\arrow[r,hook] &\OO_{\mathbb P^4}^{\oplus N}\arrow[r,two heads] &\QQ
\end{tikzcd}\ {\rm s.t.}\\
& \QQ\ {\rm of\ finite\ support\ in\ }\mathbb A^4\cong\mathbb P^4/D
\end{aligned}
\right\}\right/{\rm iso},
\end{equation}
parametrising torsion--free sheaves on $\mathbb P^4$ fitting in a given short exact sequence. Of course the previous moduli space is identified with the quot scheme of points in $\mathbb A^4$, as the Grothendieck moduli functor
\[
    \QUOT_{\mathbb P^4}(\OO^{\oplus N},k):\Sch^{\rm op}_{\mathbb C}\to\Sets
\]
contains an open subfunctor $\QUOT_{\mathbb A^4}(\OO^{\oplus N},k)\hookrightarrow\QUOT_{\mathbb P^4}(\OO^{\oplus N},k)$ parametrising precisely the quotients above. What one might want to do is then to study a model for deformations of the moduli space of sheaves
\[
\MM_{r,n}(\mathbb  P^4)=\left.\left\{
\begin{aligned}
&\EE\in\Coh(\mathbb P^4),\ \EE\ {\rm torsion-free},\ \rk(\EE)=N,\\
& c_1(\EE)=c_2(\EE)=c_3(\EE)=0,\ c_4(\EE)=k,\ \EE|_{\wp^i_\infty}\cong\OO^{\oplus N}_{\mathbb P^4}\\
&H^3(\mathbb P^4,\EE(-3))=0,\ H^2(\mathbb P^4,\EE(-\ell))=0,\forall \ell
\end{aligned}
\right\}\right/{\rm iso},
\]
where $\wp_\infty^i$, $i=1,\dots,4$, are hyperplanes at infinity in $\mathbb P^4$ defined by $z_i=0$ in homogeneous coordinates. The aim is to generalize the monad construction for the usual $\SU(N)$ instantons to the case of $\mathbb C^4$. Let then $\EE\in\Coh(\mathbb P^4)$ a sheaf on $\mathbb P^4$ as in the definition and consider $\mathbb P^4\times\mathbb P^4$ with the projections on the two factors
\[\begin{tikzcd}
 & \mathbb P^4\times\mathbb P^4\arrow[dl,"p_1"]\arrow{dr}[swap]{p_2} & \\
\mathbb P^4 & & \mathbb P^4
\end{tikzcd}\]
One then has the Koszul resolution of $\OO_\Delta$, $\Delta$ being the diagonal $\Delta\cong\mathbb P^4\hookrightarrow\mathbb P^4\times\mathbb P^4$:
\[\begin{tikzcd}
0\arrow[r] & \bigwedge^{4}\widehat\OO\arrow[r] & \bigwedge^{3}\widehat\OO\arrow[r]\arrow[d,phantom, ""{coordinate, name=Z}] & \bigwedge^{2}\widehat\OO\arrow[dlll,rounded corners,to path={ --([xshift=2ex]\tikztostart.east)|- (Z)[near end]\tikztonodes-| ([xshift=-2ex]\tikztotarget.west)-- (\tikztotarget)}]\\
\widehat\OO\arrow[r] & \OO_{\mathbb P^4\times\mathbb P^4}\arrow[r] & \OO_\Delta\arrow[r] & 0,
\end{tikzcd}\]
where $\widehat\OO=\OO_{\mathbb P^4}(1)\boxtimes\QQ^\vee$,\footnote{For any two coherent sheaves $\FF$, $\GG$ on $\mathbb P^4$, we define a sheaf on $\mathbb P^4\times\mathbb P^4$ as $\FF\boxtimes\GG=p_1^\ast\FF\otimes p_2^\ast\GG$.} and $\QQ\cong T_{\mathbb P^4}(-1)$. The previous sequence tells us that $[\OO_\Delta]\cong[\bigwedge^\bullet\left(\OO_{\mathbb P^4}(-1)\boxtimes\QQ^\vee\right)]$ in the derived category. We then define
\[
C^p=\bigwedge^{-i}\left(\OO_{\mathbb P^4}(-1)\boxtimes\QQ^\vee\right),
\]
by means of which we will define the Beilinson spectral sequence in this case. As $\EE\in\Coh(\mathbb P^4)$, we have the trivial identity $p_{1\ast}\left(p_2^\ast\EE\otimes\OO_\Delta\right)=\EE$, and if we replace $\OO_\Delta$ by its Koszul resolution we get the double complex for the hyperdirect image, which can be expressed in terms of the Fourier-Mukai transform
\[
\bold R^\bullet p_{1,\ast}\left(p_2^{\ast}\EE\otimes C^\bullet\right).
\]
There are then two different spectral sequences that can be taken for the Fourier-Mukai transform. One of them gives back the trivial identity we started with, while the other one has $E_1$--term given by
\[
E_1^{p,q}=R^qp_{1,\ast}\left(p_2^\ast\otimes C^p\right),
\]
and as $C^p=\FF_1^p\boxtimes\FF_2^p$, the $E_1$--term can be written as
\[
E_1^{p,q}=\FF_1^p\otimes H^q(\mathbb P^4,\EE\otimes\FF_2^p).
\]
This spectral sequence converges to
\[
E_\infty^{q,p}=\left\{
\begin{aligned}
& \EE(-\ell),\qquad &{\rm if\ } q+p=0\\
& 0, &{\rm otherwise}
\end{aligned}
\right.
\]
for each $\ell\ge 0$. We can actually make the first term in the sequence explicit:
\[
E_1^{p,q}=\OO_{\mathbb P^4}(p)\otimes H^q\left(\mathbb P^4,\EE(-\ell)\otimes\Omega_{\mathbb P^4}^{-p}(-p)\right),
\]
and being an object in the derived category, we can summarize it by the following complexes.
\begin{equation}\label{eqn:spectral-sequence}
\begin{aligned}
&\begin{tikzcd}
E_1^{-4,4}\arrow[r] & E_1^{-3,4}\arrow[r] & E_1^{-2,4}\arrow[r] & E_1^{-1,4}\arrow[r] & E_1^{0,4}
\end{tikzcd}\\
&\begin{tikzcd}
E_1^{-4,3}\arrow[r] & E_1^{-3,3}\arrow[r] & E_1^{-2,3}\arrow[r] & E_1^{-1,3}\arrow[r] & E_1^{0,3}
\end{tikzcd}\\
&\begin{tikzcd}
E_1^{-4,2}\arrow[r] & E_1^{-3,2}\arrow[r] & E_1^{-2,2}\arrow[r] & E_1^{-1,2}\arrow[r] & E_1^{0,2}
\end{tikzcd}\\
&\begin{tikzcd}
E_1^{-4,1}\arrow[r] & E_1^{-3,1}\arrow[r] & E_1^{-2,1}\arrow[r] & E_1^{-1,1}\arrow[r] & E_1^{0,1}
\end{tikzcd}\\
&\begin{tikzcd}
E_1^{-4,0}\arrow[r] & E_1^{-3,0}\arrow[r] & E_1^{-2,0}\arrow[r] & E_1^{-1,0}\arrow[r] & E_1^{0,0}
\end{tikzcd}\\
\end{aligned}
\end{equation}
Though the spectral sequence is naturally a doubly graded object, fixing certain conditions on the sheaf cohomology groups makes it collapse to an ordinary sequence. Indeed it is easy, albeit tedious, to show by means of homological algebra that the appropriate conditions are exactly those in the definition of $\MM_{N,k}(\mathbb P^4)$.
 By doing so, the spectral sequence displayed in \eqref{eqn:spectral-sequence} converges to $E_\infty^{-1,1}\cong\EE(-2)$, all the other terms being identically vanishing. Moreover the $E_1^{p,q}$ term of the Beilinson spectral sequence is reduced to the following \eqref{eqn:spectral-sequence-reduced}
\begin{equation}\label{eqn:spectral-sequence-reduced}
\begin{tikzcd}
0\arrow[r] & A\otimes\OO_{\mathbb P^4}(-4)\arrow[r] & B\otimes\OO_{\mathbb P^4}(-3)\arrow[r]\arrow[d,phantom, ""{coordinate, name=Z}] & C\otimes\OO_{\mathbb P^4}(-2)\arrow[dll,rounded corners,to path={ --([xshift=2ex]\tikztostart.east)|- (Z)[near end]\tikztonodes-| ([xshift=-2ex]\tikztotarget.west)-- (\tikztotarget)}]
\\
& D\otimes\OO_{\mathbb P^4}(-1)\arrow[r] & E\otimes\OO_{\mathbb P^4}\arrow[r] & 0,
\end{tikzcd}
\end{equation}
where
\begin{align*}
    &A=H^1(\mathbb P^4,\EE(-3)), & B=H^1(\mathbb P^4,\EE(-2)\otimes\Omega_{\mathbb P^4}^3(3)),\\
    & C=H^1(\mathbb P^4,\EE(-2)\otimes\Omega_{\mathbb P^4}^2(2)), & D=H^1(\mathbb P^4,\EE(-2)\otimes\Omega_{\mathbb P^4}^1(1)),\\
    &E=H^1(\mathbb P^4,\EE(-2)).
\end{align*}
We then see that \eqref{eqn:spectral-sequence-reduced} is a perfect extended monad in the sense of \cite[Definition 3.1/3.2]{Henni2017}, and it provides a model for the deformation of the Quot scheme of points on $\mathbb A^4$. Analogously one can lift the set theoretic isomorphism between moduli spaces of framed sheaves and quot schemes of points to a scheme theoretic isomorphism by virtue of a version in families of Beilinson's theorem. Indeed, let $S$ be a scheme over $\mathbb C$. Given any coherent sheaf $\EE$ on $\mathbb P^m\times S$ there exists a spectral sequence $E_i^{p,q}$ whose $E_1^{p,q}$ term is
\[
    E_1^{p,q} = \OO_{\mathbb P^m}(p)\boxtimes R^qp_{2,\ast}\left(\EE\otimes\Omega^{-p}_{\mathbb P^4\times S/S}(-p)\right),
\]
where $p_2$ is the projection of $\mathbb P^m\times S$ on the second factor. The spectral sequence $E_i^{p,q}$ converges to
\[
E_\infty^{q,p}=\left\{
\begin{aligned}
& \EE,\qquad &{\rm if\ } q+p=0\\
& 0, &{\rm otherwise.}
\end{aligned}
\right.
\]

This approach was studied in much greater generality in \cite{Henni2017}, and more details about the proof can be found also in \cite{Henni2018,Henni2015}. Moreover, the scheme-theoretic isomorphism between moduli spaces of framed torsion-free sheaves on $\mathbb P^m$ and quot schemes of points on $\mathbb A^m$ was recently proved by means of an infinitesimal argument in \cite{Cazzaniga2020}.

\begin{remark}
The choice of conditions constraining the sheaves $\EE\in\Coh_{\mathbb P^4}$ are different from the instanton sheaf conditions found in \cite{Jardim2006} as the latter would exclude certain sheaf configurations, such as ideal sheaves of points, which are instead known to be interesting to the problem at hand.
\end{remark}

\subsection{Orbifold instantons and ADHM data decomposition}\label{appendix:orbifold beilinson}
It is interesting to notice here that in general it is not known whether a Gorenstein orbifold in dimension 4 admits a crepant resolution, and many singularities are terminal. Nonetheless some classification results, though a bit scattered, are available. It is known, for example, that orbifolds of the form $\mathbb C^r/G$ admit crepant resolutions in the case of finite {\it abelian} subgroups $G\subset \SL(r,\mathbb C)$ for which $\mathbb C^r/G$ is a complete intersection, \cite{Dais1998,Dais2001}, while some arithmetic condition have been derived in the case of some series of cyclic quotient singularities \cite{Dais98,Dais2006}. Moreover, contrary to the lower dimensional cases, here the Hilbert-Chow morphism
\[
  \Hilb^G(\mathbb C^r)\xrightarrow{\quad\pi\quad}\mathbb C^r/G
\]
does not necessarily provide a crepant resolution, as there are even cases in which $\Hilb^G(\mathbb C^r)$ is singular despite $\mathbb C^r/G$ being known to have projective crepant resolutions. Here $\Hilb^G(\mathbb C^r)$ denotes the $G$-Hilbert scheme of zero-dimensional $G$-invariant subschemes $Z\subset\mathbb C^r$ of length $|G|$ such that $H^0(\OO_Z)$ is the regular representation of $G$. It turns out, however, that in some cases which will be of interest to us a canonical crepant (projective) resolution of the orbifold singularity is indeed provided by the $G$-Hilbert scheme, and toric geometry techniques are well suited to check whether this is the case or not, \cite{sato2010,satoy2019}. Let's assume for the moment that $X=\Hilb^G(\mathbb C^4)$ is a crepant resolution of $\mathbb C^4/G$: this will enable us to justify the ADHM data decomposition which we will use in the following in order to study the orbifold instantons in eight dimensions. Take then the universal object $\mathcal Z\subset X\times\mathbb C^4$ with the projections $\pi_i$ on the $i$-th factor.
\[\begin{tikzcd}
 & \mathcal Z\arrow[d,dash]\\
 & X\times\mathbb C^4\arrow[dl,"\pi_1"]\arrow{dr}[swap]{\pi_2} & \\
X & & \mathbb C^4
\end{tikzcd}\]
We can then introducte the tautological bundle $\TT$ on $X$ by pushing forward
\[
  \TT=\pi_{1\ast}\OO_{\mathcal Z}.
\]
Under the $G$-action, $\TT$ transforms under the regular representation, and it is easy to show that its fibers are of the form $\mathbb C[z_1,z_2,z_3,z_4]/I\cong H^0(\OO_Y)$, where $I$ is a $G$-invariant ideal corresponding to the zero-dimensional subscheme $Y\subseteq\mathbb C^4$ of length $|G|$. Multiplication by the coordinates along the fibers of $\TT$ induces a $G$-invariant homomorphism $B\in Q\otimes\End(\TT)$ (where $Q$ denotes the regular representation of $G$), which is representable by a quadruple of endomorphisms $(B_1,\dots,B_4)$, such that $B\wedge B=\sum_{a<b}[B_a,B_b]=0\in\Hom_G(\TT,\bigwedge^2Q\otimes\TT)$. As we noticed in section \ref{sec:adhm-data-decomposition} the regular representation $Q$ may be decomposed in irreducible representations of $G$. This induces a decomposition of the tautological bundle as
\[
    \TT=\bigoplus_{r}\TT_r\otimes\rho_r.
\]
The monad construction for the ADHM representation of orbifold instantons then follows from a generalisation of \cite{Kronheimer1990,Nakajima2007,Cirafici2011}. One starts from the resolution of the diagonal in $X\times X$, which is
\[\begin{tikzcd}[column sep = small]
\left(\TT\boxtimes\TT^\vee\otimes\bigwedge^4 Q^\vee\right)^G\arrow[r] & \left(\TT\boxtimes\TT^\vee\otimes\bigwedge^3 Q^\vee\right)^G\arrow[r]\arrow[d,phantom, ""{coordinate, name=Z}] & \left(\TT\boxtimes\TT^\vee\otimes\bigwedge^2 Q^\vee\right)^G\arrow[dll,rounded corners,to path={ --([xshift=2ex]\tikztostart.east)|- (Z)[near end]\tikztonodes-| ([xshift=-2ex]\tikztotarget.west)-- (\tikztotarget)}]\\
\left(\TT\boxtimes\TT^\vee\otimes Q^\vee\right)^G\arrow[r] & \left(\TT\boxtimes\TT^\vee\right)^G\arrow[r] & \OO_\Delta.
\end{tikzcd}\]

We then compactify $X$ to $\overline X$ by compactifying $\mathbb C^4/G$ to $\mathbb P^4/G$ and resolving the singularity at the origin. We can do this by defining $\overline X=X\sqcup\PP_\infty$, where $\PP_\infty\cong\mathbb P^3/G$. This is useful as we can then glue objects on $X$ and $G$-invariant objects on $\PP_\infty$ so as to get globally defined objects on $\overline X$. A globally defined resolution of the diagonal sheaf in $\overline X\times\overline X$ can then be obtained in this way, and we get
\[\begin{tikzcd}[column sep = small]
0\arrow[r] & \left(\TT(-4\PP_\infty)\boxtimes\TT^\vee\otimes\bigwedge^4 \QQ^\vee\right)^G\arrow[r]\arrow[d,phantom, ""{coordinate, name=Z}] & \left(\TT(-3\PP_\infty)\boxtimes\TT^\vee\otimes\bigwedge^3 \QQ^\vee\right)^G\arrow[dl,rounded corners,to path={ --([xshift=2ex]\tikztostart.east)|- (Z)[near end]\tikztonodes-| ([xshift=-2ex]\tikztotarget.west)-- (\tikztotarget)}]\\
 & \left(\TT(-2\PP_\infty)\boxtimes\TT^\vee\otimes\bigwedge^2 \QQ^\vee\right)^G\arrow[r]\arrow[d,phantom, ""{coordinate, name=Z}] & \left(\TT(-\PP_\infty)\boxtimes\TT^\vee\otimes \QQ^\vee\right)^G\arrow[dl,rounded corners,to path={ --([xshift=2ex]\tikztostart.east)|- (Z)[near end]\tikztonodes-| ([xshift=-2ex]\tikztotarget.west)-- (\tikztotarget)}]\\
 & \left(\TT\boxtimes\TT^\vee\right)^G\arrow[r] & \OO_\Delta\arrow[r] & 0.
\end{tikzcd}\]

The construction then follows the same strategy adopted in the previous section. In particular, if $\EE$ is a coherent sheaf on $\overline X$, one can take its Fourier-Mukai transform, where now $C^\bullet$ is the resolution of the diagonal $\OO_\Delta$ in $\overline X$. Then, associated to the sheaf $\EE(-\ell)=\EE\otimes\OO_{\overline X}(-\ell\PP_\infty)$ there is a spectral sequence $E_i^{p,q}$, whose $E_1^{p,q}$-term is given by
\[
    E_1^{p,q}=\left(\TT(p)\otimes H^q(\overline X,\EE(-\ell)\times\TT^\vee\otimes\bigwedge^{-p}\QQ^\vee)\right)^G.
\]
The tensor product decomposition of the tautological bundle $\TT$ as $\TT=\bigoplus\TT_r\otimes\rho_r$ makes it possible to reduce the homological algebra coming from the orbifold spectral sequence to the easier $\mathbb P^4$ case. Indeed, cohomology groups relevant to the computations all take the form $H^q(\overline X,\EE(-\ell)\otimes\TT_r)$, which can in turn be interpreted as the equivariant decomposition of the ADHM datum.

\bibliographystyle{biblio}
\bibliography{bib.bib}

\end{document}

%% file: macros.tex
\usepackage{amscd,mathrsfs}
\usepackage{amsfonts,mathdots,dsfont}
\usepackage{psfrag,slashed}

\makeatletter
\tikzset{join/.code=\tikzset{after node path={%
\ifx\tikzchainprevious\pgfutil@empty\else(\tikzchainprevious)%
edge[every join]#1(\tikzchaincurrent)\fi}}}
\makeatother

\tikzset{>=stealth',every on chain/.append style={join},
         every join/.style={->}}
\tikzset{
    >=stealth',
    punkt/.style={
           rectangle,
           rounded corners,
           draw=black, very thick,
           text width=6.5em,
           minimum height=2em,
           text centered},
    pil/.style={
           ->,
           thick,
           shorten <=2pt,
           shorten >=2pt,}
}

\newcommand{\bea}{\begin{eqnarray}}
\newcommand{\eea}{\end{eqnarray}}
\newcommand{\be}{\begin{equation}}
\newcommand{\ee}{\end{equation}}

\def\i{{\iota}}

\DeclareMathAlphabet{\mathpzc}{OT1}{pzc}{m}{it}